\documentclass[12pt,journal,draftclsnofoot,onecolumn]{IEEEtran}

\usepackage{amsmath,amsfonts,amssymb}
\usepackage{dsfont}
\usepackage[hidelinks]{hyperref}
\usepackage{mathenv}
\usepackage{mathtools}
\usepackage[noadjust]{cite}
\usepackage{graphicx}
\usepackage{subcaption}
\usepackage{booktabs}

\newenvironment{proofof}[1]{\begin{IEEEproof}[Proof of #1]}{\end{IEEEproof}}
\newenvironment{sketch}{\begin{IEEEproof}[Sketch of Proof]}{\end{IEEEproof}}
\newenvironment{verificationOf}[1]{\begin{IEEEproof}[Verification of #1]}{\end{IEEEproof}}
\newcommand{\proofref}[2][Proof]{\leavevmode\unskip\hfill\penalty0\hbox{\quad}\nolinebreak\hfill\hbox{(#1 in #2.)}}
\newcommand{\verificationref}[1]{\proofref[Verification]{#1}}

\newtheorem{theorem}{Theorem}[section]
\newtheorem{corollary}[theorem]{Corollary}
\newtheorem{lemma}[theorem]{Lemma}
\newtheorem{proposition}[theorem]{Proposition}
\newtheorem{remark}{Remark}[section]

\newtheorem{conjecture}{Conjecture}[section]

\newcommand{\diff}{\mathrm{d}}
\newcommand{\eqdef}{:=}
\newcommand{\expect}{\mathbb{E}}
\newcommand{\en}{\mathrm{e}}
\newcommand{\real}{\mathbb{R}}
\newcommand{\nnreal}{\real_{\ge 0}}
\newcommand{\relvar}[2]{\buildrel\textrm{#2}\over{#1}}
\newcommand{\eqvar}[1]{\relvar{=}{#1}}
\newcommand{\gevar}[1]{\relvar{\ge}{#1}}
\newcommand{\levar}[1]{\relvar{\le}{#1}}

\newcommand{\indicator}{\mathds{1}}

\newcommand{\greedy}{s_{\text{grd}}}
\newcommand{\fixedFraction}{s_{\text{ff}}}
\newcommand{\maximinOptimalLinear}{s_{\text{mol}}}
\newcommand{\maximinOptimalLinearHat}{\hat{s}_{\text{mol}}}
\newcommand{\camfOptimalLinear}{s_{\times}}
\newcommand{\camfOptimalLinearHat}{\hat{s}_{\times}}
\newcommand{\caagOptimalLinear}{s_{+}}

\newcommand{\parameterizedPolicy}[3][]{%
  #2_{\text{#3}%
  \ifx%
    \\#1\\%
  \else%
    (#1)%
  \fi}%
}
\newcommand{\maximinOptimalPolicy}[1][]{\parameterizedPolicy[#1]{\sigma}{mo}}
\newcommand{\robustClippedAffinePolicy}[1][]{\parameterizedPolicy[#1]{\sigma}{rca}}
\newcommand{\rcaOL}[1][]{\parameterizedPolicy[#1]{\sigma}{rca-ol}}
\newcommand{\rcaMFOL}[1][]{\parameterizedPolicy[#1]{\sigma}{rca-mfol}}
\newcommand{\rcaAGOL}[1][]{\parameterizedPolicy[#1]{\sigma}{rca-agol}}
\newcommand{\lyapunovPolicy}{\sigma_{\text{lyap}}}

\DeclareMathOperator{\mcr}{MCR}

\newcommand{\bernoulli}{\tilde{\mathrm{B}}}

\begin{document}

\title{On Linear Power Control Policies for Energy Harvesting Communications}

\author{
  Hafez~M.~Garmaroudi,
  Zikai~Dou,
  Shengtian~Yang,~\IEEEmembership{Senior Member,~IEEE},
  and~Jun~Chen,~\IEEEmembership{Senior Member,~IEEE}%
  \thanks{Corresponding author: Shengtian~Yang.}
  \thanks{H.~M.~Garmaroudi, Z.~Dou, and J.~Chen are with the Department of Electrical and Computer Engineering, McMaster University, Hamilton, ON L8S 4K1, Canada (e-mail: \{mousas15, douz7, chenjun\}@mcmaster.ca).}%
  \thanks{S.~Yang is with the School of Information and Electronic Engineering (Sussex Artificial Intelligence Institute), Zhejiang Gongshang University, Hangzhou 310018, China (e-mail: \mbox{yangst@codlab.net}).}%
}

\maketitle

\begin{abstract}
  This paper studies optimal linear power control for battery-limited energy harvesting communications.
  It provides a systematic analysis of linear power control policies, covering the greedy and fixed-fraction policies as special cases.
  Three optimality notions are introduced: the maximin optimal linear policy for a given battery capacity and mean-to-capacity ratio (MCR), and two capacity-agnostic policies that minimize the nominal additive gap and maximize the nominal multiplicative factor, respectively.
  Except the capacity-agnostic additive-gap optimal linear policy, which coincides with the fixed-fraction policy, the other two optimal linear policies are novel and constitute the main contributions of this paper.
  It is shown, among others, that the worst nominal multiplicative factor for both novel policies is approximately $0.6530$, a substantial improvement over the fixed-fraction policy's value of $0.5$.
  Simulations show that under quasi-static fading, the maximin optimal linear policy performs comparable to the maximin optimal policy (the top-performing policy), while the capacity-agnostic multiplicative-factor optimal linear policy performs slightly worse; nevertheless, both novel policies significantly outperform the fixed-fraction policy in the low-to-medium signal-to-noise ratio (SNR) regime.
  Moreover, this paper also investigates the optimality of the greedy policy for certain families of energy-arrival distributions, and establishes the tightest semi-universal bounds on the battery-capacity threshold for greedy optimality.
\end{abstract}


\section{Introduction}

Due to recent advances in Internet of Things, wireless nodes have become vital as they provide accessibility to distant locations or provide sensor measurements for different applications.
Seeking greater mobility and flexibility, most of these wireless nodes rely on batteries for their operation, instead of resorting to the power line.
The ability to harvest energy from the environment significantly increases the lifespan of wireless nodes and enhances their independence, self-reliance, and self-sustainability.
A particular problem for these energy harvesting communication systems is to find the optimal policy for energy expenditure that maximizes the long-term average throughput.
This problem has been studied intensely in recent years
\cite{%
  sharma2010,
  ozel2011, yang2012, tutun2012, ho2012,
  ozel2012, blasco2013, wang2013, srivasta2013, khuzani2014, xu2014, rajesh2014,
  ulkus2015,
  dong2015, amirnavaei2016online, shaviv2016universally, shaviv2017approximately, arafa2018fixed,
  yang2020.6maximin, yang2020maximin, wang2021optimality, zibaeenejad2022optimal, yang2025power
}
with particular attention to two different settings: offline power control and online power control.

In the offline setting, the energy-arrival process is known in advance, so the underlying distribution does not have much relevance as far as the design of power control policy is concerned.
The optimal offline policy admits a relatively simple characterization, which basically strives to keep the battery level at a fixed value while avoiding overflows \cite{yang2012, tutun2012, ozel2011}.

By contrast, in the online setting, the nodes do not know the realization of the energy-arrival process ahead of time. As such, the distribution of energy arrivals has to be taken into account when it comes to policy design.
In general, the optimal online (power control) policy is only implicitly characterized via the Bellman equation.
One exception is the Bernoulli energy-arrival case, for which the optimal online policy is known explicitly \cite{shaviv2016universally, yang2020.6maximin, yang2020maximin}.

In view of the difficulty in finding the optimal online policy and its potential high complexity, some efforts have been made to analyze the performance of certain simple policies. The greedy policy,  which depletes the battery in every time slot,
is thoroughly investigated in \cite{wang2021optimality}, which reveals that this seemingly trivial policy is actually optimal in the low-battery-capacity regime. Also noteworthy is the fixed-fraction policy introduced in \cite{shaviv2016universally}.
This policy expends a constant fraction $p$ of the available energy in each time slot, where $p$ is the mean-to-capacity ratio (MCR), a key statistic measuring the average energy availability relative to the battery capacity.
Despite its simplicity, the fixed-fraction policy enjoys the remarkable property that its performance is universally near optimal in terms of additive and multiplicative gaps from the fundamental limit.

Interestingly, both the greedy policy and the fixed-fraction policy are linear policies in the sense that the amount of energy expended in each time slot is a time-invariant linear function of the battery level. They only differ in the slopes of their respective linear functions ($1$ for the greedy policy and $p$ for the fixed-fraction policy).
The results in \cite{wang2021optimality, shaviv2016universally} suggest that in addition to having the obvious advantage of low implementation complexity, linear policies can be performance-wise quite competitive. Motivated by this observation, we attempt to conduct a systematic study of such policies in the present work.
The main contributions of this paper are as follows.

\begin{itemize}
  \item We present three notions of optimal linear policies: the maximin optimal linear policy $\maximinOptimalLinear$, the capacity-agnostic additive-gap optimal linear policy $\caagOptimalLinear$, and the capacity-agnostic multiplicative-factor optimal linear policy $\camfOptimalLinear$.
  Except for the second, which coincides with the fixed-fraction policy, the other two are novel.
  We run comprehensive simulations comparing these linear policies with the optimal online policy and representative nonlinear baselines under quasi-static fading and block-fading channels.
  Under quasi-static fading, the maximin optimal linear policy $\maximinOptimalLinear$ closely matches the performance of the maximin optimal policy $\maximinOptimalPolicy$ (the top-performing policy), incurring an average loss below $1\%$ and a worst-case loss of about $2\%$ (relative to the optimal online policy).
  The capacity-agnostic multiplicative-factor optimal linear policy $\camfOptimalLinear$ performs slightly worse than $\maximinOptimalLinear$, but remains significantly better than the capacity-agnostic additive-gap optimal linear policy $\caagOptimalLinear$, which performs well only in the high signal-to-noise ratio (SNR) regime.
  Although policies not using (instantaneous) channel-power-gain information, including linear policies, degrade markedly in the low-to-medium SNR regime under block fading, two robust clipped affine (RCA) policies derived from $\maximinOptimalLinear$ and $\camfOptimalLinear$ (see~\cite{wu2026clipped} for the general RCA design) exploit channel-power-gain information effectively and deliver the best performance: average losses below $1.5\%$ and worst-case losses below $4\%$.

  \item We perform a systematic analysis of the nominal additive gap and nominal multiplicative factor for the proposed optimal linear policies (see Table~\ref{tab:linear_policy_performance}).
  In particular, we show that the worst nominal multiplicative factor for both $\maximinOptimalLinear$ and $\camfOptimalLinear$ is approximately $0.6530$, a substantial improvement over the fixed-fraction policy's value of $0.5$.
  We also characterize two fundamental saddle-point structures:
  the nominal additive gap and the nominal multiplicative factor each induce a saddle-point structure, and the policies $\caagOptimalLinear$ and $\camfOptimalLinear$ constitute the corresponding saddle points.

  \item We investigate the optimality of the greedy policy for certain families of energy-arrival distributions.
  We establish the tightest semi-universal bounds on the battery-capacity threshold for greedy optimality (hereafter the greedy threshold), thereby confirming the tightness of the bounds in~\cite[Props.~4 and 5]{wang2021optimality}.
  We also derive tight semi-universal bounds on the greedy threshold for clipped energy-arrival distributions.
  In particular, we present a tight upper bound on the greedy threshold in terms of the MCR (see Table~\ref{tab:linear_policy_performance}).
\end{itemize}

\begin{table}
  \caption{The Performance of Worst-Case Optimal Linear Policies and Greedy Policy}
  \label{tab:linear_policy_performance}
  \centering
  \begin{tabular}{cccc}
    \toprule
    Type of linear policies                         & Slope                                                               & Additive gap (nat)                     & Multiplicative factor                  \\
    \midrule
    maximin optimal                                 & $\maximinOptimalLinear(c,p)$                                        & $\le\frac{1}{2}$                       & $\gtrapprox 0.6530$                    \\
    & (Eqs.~\eqref{eq:mol_definition} and~\eqref{eq:mol_approximation})   & (Thm.~\ref{th:mol_worst_performance})  & (Thm.~\ref{th:mol_worst_performance})  \\
    capacity-agnostic additive-gap optimal          & $\caagOptimalLinear(p)=p$                                           & $\le\frac{1}{2}$                        & $\ge\frac{1}{2}$ \\
    & (Eq.~\eqref{eq:agol_definition} and Thm.~\ref{th:ag_minimax})       & (Thm.~\ref{th:agol_worst_performance}) & (Thm.~\ref{th:agol_worst_performance}) \\
    capacity-agnostic multiplicative-factor optimal & $\camfOptimalLinear(p)$                                             & $\lessapprox 0.7292$           & $\gtrapprox 0.6530$ \\
    & (Eqs.~\eqref{eq:mfol_definition} and~\eqref{eq:mfol_approximation}) & (Thm.~\ref{th:mfol_worst_performance}) & (Thm.~\ref{th:mfol_worst_performance}) \\
    greedy                                          & $1$                                                                 & $>0$ if $c>\overline{c}''(p)$          & $<1$ if $c>\overline{c}''(p)$          \\
    &                                                                     & (Thm.~\ref{upper_c_for_mcr})           & (Thm.~\ref{upper_c_for_mcr})           \\
    \bottomrule
  \end{tabular}
\end{table}

The rest of the paper is organized as follows.
Section~\ref{sec:formulation} introduces the problem formulation and notation.
Section~\ref{sec:worst_case_optimal_linear_policies} presents three notions of optimal linear policies and their simulation-based evaluation.
Section~\ref{sec:performance_analysis} analyzes the performance of the proposed linear policies and the greedy policy.
Section~\ref{sec:conclusion} concludes the paper, and the appendices provide the proofs and numerical verification of all nontrivial results.
For convenience, common notation used throughout the paper is summarized in Table~\ref{tab:notation} (see Table~\ref{tab:linear_policy_performance} for linear‑policy performance and notation).

\begin{table}
  \centering
  \caption{Common Notation in This Paper}
  \label{tab:notation}
  \begin{tabular}{ll}
    \toprule
    Symbol                      & Description                                                                  \\
    \midrule
    $c$                         & Battery capacity                                                             \\
    $Q$                         & Marginal distribution of i.i.d.\ energy-arrival process                      \\
    $\tilde{\mathrm{B}}_{c,p}$  & Bernoulli distribution (Eq.~\eqref{eq:bernoulli})                            \\
    $\mu$                       & Mean of $Q$                                                                  \\
    $\bar\mu$                   & Clipped mean of $Q$ (Eq.~\eqref{eq:clipped_mean})                            \\
    $p=\mathrm{MCR}_c(Q)$       & Mean-to-capacity ratio (MCR) (Eq.~\eqref{eq:mcr})                            \\
    $r(x,\gamma)$               & Reward function (Eq.~\eqref{eq:awgn})                                        \\
    $G_Q(c,\sigma)$             & Additive gap (Eq.~\eqref{eq:additive_gap})                                   \\
    $F_Q(c,\sigma)$             & Multiplicative factor (Eq.~\eqref{eq:multiplicative_factor})                 \\
    $\underline{\Gamma}(c,p,s)$ & Worst-case (Bernoulli) throughput (Eq.~\eqref{lower_gamma})                  \\
    $\overline{G}(c,p,s)$       & Nominal additive gap (Eq.~\eqref{eq:nominal_additive_gap}                    \\
    $\underline{F}(c,p,s)$      & Nominal multiplicative factor (Eq.~\eqref{eq:nominal_multiplicative_factor}) \\
    \bottomrule
  \end{tabular}
\end{table}

\section{Problem Formulation}
\label{sec:formulation}

Consider a discrete-time energy harvesting communication system with a battery of capacity $c$.
The amount of energy harvested at time $t$ is denoted by $E_t$.
The process $E^\infty=(E_t)_{t=1}^\infty$ of energy arrivals is assumed to be independent and identically distributed (i.i.d.) with marginal distribution $Q$, a probability measure on $\nnreal$ (with the associated Borel $\sigma$-field).
Under the assumption of the harvest-store-use architecture, the harvested energy is first stored in the battery and then consumed to transmit data over a point-to-point quasi-static fading AWGN channel.
Let $B_t$ be the battery level at time $t$ (after the arrival of $E_t$) and $G_t$ the consumed energy in time slot $t$.
Then
\[
  B_{t+1}
  = \min\{B_t - G_t + E_{t+1}, c\}
\]
with the admissibility condition $G_t\le B_t$ for all $t\ge 1$.

In its most general form, $G_t$ is a function of $E^t=(E_1,\ldots,E_t)$, that is, $G_t=\pi_t(E^t)$. A sequence $\pi^\infty=(\pi_t)_{t=1}^\infty$ of such functions forms an (admissible) online power control policy.
The induced (long-term average) throughput of the system is defined as
\begin{equation}
  \mathcal{T}_c(\pi^\infty,Q)
  \eqdef \liminf_{n\to\infty} \frac{1}{n} \expect\left(\sum_{t=1}^n r(G_t, \gamma)\right),\label{throughput}
\end{equation}
where
\begin{equation}
  r(x,\gamma)
  \eqdef \frac{1}{2}\log(1+\gamma x) \quad \text{(nats/real channel use)} \label{eq:awgn}
\end{equation}
is the capacity of the quasi-static fading AWGN channel with $\gamma$ being the channel SNR coefficient that remains constant throughout the entire transmission time.
Throughout this paper, $\log$ denotes the natural logarithm (base $\mathrm{e}$).

\begin{remark}\label{snr_normalization_remark}
  With no loss of generality, we assume that the channel SNR coefficient $\gamma=1$, as it can be absorbed into the definitions of $x$ and all other energy quantities.
  Consequently, all these quantities, in particular the battery capacity $c$ in this paper, should be understood as being on receiver side and will change as the channel SNR coefficient varies.
\end{remark}

The following quantities will be used frequently in our analysis.
The mean and the clipped mean (also called effective mean) of $Q$ are denoted by
\begin{equation}
  \mu = \expect_{X\sim Q} X \label{eq:mean}
\end{equation}
and
\begin{equation}
  \bar{\mu} = \expect_{X\sim Q}(\min\{X,c\}),\label{eq:clipped_mean}
\end{equation}
respectively.
The mean-to-capacity ratio (MCR) $p$ of $Q$ is defined by
\begin{equation}
  p
  = \mcr_c(Q)
  \eqdef \frac{\bar{\mu}}{c}
  \quad\text{\cite[Def.~3]{yang2020maximin}}.\label{eq:mcr}
\end{equation}

The core of the problem is to find an optimal (online power control) policy to achieve the maximum throughput
\[
  \mathcal{T}_c^*(Q)
  \eqdef \sup_{\pi^\infty} \mathcal{T}_c(\pi^\infty,Q),
\]
where the supremum is taken over all policies.
In general, under certain conditions (e.g., \cite[Thm.~6.1]{arapostathis1993discrete}), there exists a stationary policy (i.e., a time-invariant policy depending on $E^t$ only through $B_t$), a mapping $\sigma: [0,c]\to [0,c]$ satisfying $\sigma(x)\le x$ such that $(G_t=\sigma(B_t))_{t=1}^\infty$ achieves the maximum throughput.
In the sequel, we restrict attention to stationary policies, particularly linear ones.
Given a stationary policy $\sigma$, its performance can be evaluated using two metrics:
\begin{enumerate}
  \item Additive gap:
  \begin{equation}
    G_Q(c,\sigma)
    \eqdef \mathcal{T}_c^*(Q) - \mathcal{T}_c(\sigma,Q);\label{eq:additive_gap}
  \end{equation}
  \item Multiplicative factor:
  \begin{equation}
    F_Q(c,\sigma)
    \eqdef \frac{\mathcal{T}_c(\sigma,Q)}{\mathcal{T}_c^*(Q)}.\label{eq:multiplicative_factor}
  \end{equation}
\end{enumerate}
For a linear policy of the form $\varphi_s(b)\eqdef sb$, where the slope $s\in[0,1]$ uniquely specifies $\varphi_s$, we write $F_Q(c,s)$ and $G_Q(c,s)$ for the above two metrics when no ambiguity arises.

\section{Worst-Case Optimal Linear Policies}
\label{sec:worst_case_optimal_linear_policies}

In practice, the energy-arrival distribution $Q$ is often only partially known, and even when $Q$ is specified, computing an optimal online power control policy can be computationally demanding.
Motivated by the worst-case analysis developed in~\cite{shaviv2016universally, yang2020maximin}, we derive in this section explicit rules for selecting linear policies by optimizing one or both of two performance metrics, which depend only on limited system information: the battery capacity $c$ and the MCR $p$.

It is known from~\cite[Prop.~5]{shaviv2016universally} that for any given $c$ and $p$, the worst-case distribution for a linear policy is the Bernoulli distribution
\begin{equation}
  \bernoulli_{c,p}
  \eqdef (1-p)\delta_0+p\delta_{c}\label{eq:bernoulli}
\end{equation}
with $\delta_x(A) \eqdef \indicator\{x\in A\}$.
In this case, the throughput achieved by a linear policy of slope $s$ is
\begin{equation}
  \underline{\Gamma}(c,p,s)
  \eqdef \sum_{i=0}^\infty p(1-p)^{i} r(cs(1-s)^{i})\label{lower_gamma} \quad\text{(\cite[Lemma~2]{yang2020maximin})}.
\end{equation}
On the other hand, an upper bound on the maximum throughput for any distribution with MCR $p$ is given by
\begin{equation}
  \overline{\Gamma}(c,p)
  \eqdef r(pc)
  \quad \text{(\cite[Prop.~2]{shaviv2016universally})}.\label{upper_gamma}
\end{equation}
These two bounds then lead to two practical performance metrics for a linear policy of slope $s$:
\begin{enumerate}
  \item Nominal additive gap:
  \begin{equation}
    \overline{G}(c,p,s)
    \eqdef \overline{\Gamma}(c,p) - \underline{\Gamma}(c,p,s);\label{eq:nominal_additive_gap}
  \end{equation}
  \item Nominal multiplicative factor:
  \begin{equation}
    \underline{F}(c,p,s)
    \eqdef \frac{\underline{\Gamma}(c,p,s)}{\overline{\Gamma}(c,p)}.\label{eq:nominal_multiplicative_factor}
  \end{equation}
\end{enumerate}
These quantities serve as an upper bound on $G_Q(c,s)$ and a lower bound on $F_Q(c,s)$, respectively, for any $Q$ with $\mcr_c(Q)=p$.

Under these two practical metrics, the central problem of this section is to select the slope $s$, that is, to specify a rule mapping $(c,p)$ to $s$.
The fixed-fraction and greedy policies correspond to the slopes
\begin{equation}
  \fixedFraction(p) \eqdef p
  \quad\text{and}\quad
  \greedy \eqdef 1,
\end{equation}
respectively.
Note that the fixed-fraction policy depends only on the MCR $p$, whereas the greedy policy is independent of both $c$ and $p$.

\subsection{Maximin Optimal Linear Policy for Fixed Battery Capacity and MCR}\label{subsec:maximin_optimal_linear_policy}

When both the battery capacity $c$ and the MCR $p$ are fixed, optimizing either the nominal additive gap or the nominal multiplicative factor reduces to maximizing the worst-case throughput $\underline{\Gamma}(c,p,s)$ over all linear policies.
Thus, we have the notion of \emph{maximin optimal linear policy} defined by the slope
\begin{equation}
  \maximinOptimalLinear(c,p)
  \eqdef \arg\max_{s\in [0,1]} \underline{\Gamma}(c,p,s).\label{eq:mol_definition}
\end{equation}

\begin{proposition}\label{pr:mol.special}
  If $c\leq p/(1-p)$, then $\maximinOptimalLinear(c,p)=1$.
\end{proposition}

This is a straightforward consequence of~\cite[Thm.~1]{wang2021optimality}, which shows that the greedy policy is optimal (among all online policies) for $Q=\bernoulli_{c,p}$ with $c\le p/(1-p)$.
For $c > p/(1-p)$, no closed-form expression is known for $\maximinOptimalLinear(c,p)$.
A practical way is to precompute its values on a grid and store them in a lookup table for online use.
According to Conjecture~\ref{cj:throughput_quasi_concave_in_s} (supported by extensive numerical verification), the maximizer in~\eqref{eq:mol_definition} is unique and can be found efficiently by a scalar search (e.g., golden-section search) over $s$.
Alternatively, we can use the following approximation:
\begin{equation}
  \maximinOptimalLinearHat(c,p)
  \eqdef \frac{\maximinOptimalPolicy[p](c)}{c},\label{eq:mol_approximation}
\end{equation}
where $\maximinOptimalPolicy[p]$ denotes the maximin optimal policy in~\cite[Thm.~1]{yang2020maximin}, which is one of the optimal stationary policies for energy-arrival distribution $\bernoulli_{c,p}$.
Numerical results show that
\(
% eh4-16
-0.086
\le \maximinOptimalLinearHat(c,p) - \maximinOptimalLinear(c,p)
\le 0
\)
for all $c>0$ and $p\in(0,1)$.

Thanks to~\cite[Lemma~2]{yang2020maximin}, the throughput of the maximin optimal policy under Bernoulli energy arrivals can be evaluated via an expression analogous to~\eqref{lower_gamma}.
This enables efficient numerical evaluation of the multiplicative factor and additive gap of the maximin optimal linear policy under bernoulli energy arrivals.
We have
\begin{gather}
  % eh4-15
  \sup_{c>0, p\in(0,1)} G_{\bernoulli_{c,p}}(c, \maximinOptimalLinear(c,p))
  \approx 0.0061,\label{eq:mol_additive_gap}\\
  \inf_{c>0, p\in(0,1)} F_{\bernoulli_{c,p}}(c, \maximinOptimalLinear(c,p))
  \approx 0.9855.\label{eq:mol_multiplicative_factor}
\end{gather}

\subsection{Capacity-Agnostic Optimal Linear Policies for Fixed MCR}\label{subsec:capacity_agnostic_optimal_linear_policies}

As noted in Remark~\ref{snr_normalization_remark}, with $\gamma=1$, all energy quantities should be understood as scaled by the actual channel SNR coefficient, say $\gamma'$.
When $\gamma'$ is unknown or unreliable, it is desirable to use universally good policies independent of $\gamma'$.
A linear policy with fixed slope is form-invariant with respect to $\gamma'$; consequently, its performance depends on $\gamma'$ only through the capacity $c$ (which absorbs $\gamma'$).
Then it suffices to select a slope that optimizes worst-case performance over all $c>0$, which motivates the capacity-agnostic optimal linear policies defined below.
\begin{enumerate}
  \item \emph{Capacity-agnostic additive-gap optimal linear policy}:
  \begin{equation}
    \caagOptimalLinear(p)
    \eqdef \arg\min_{s\in [0,1]} \sup_{c>0} \overline{G}(c,p,s).\label{eq:agol_definition}
  \end{equation}
  \item \emph{Capacity-agnostic multiplicative-factor optimal linear policy}:
  \begin{equation}
    \camfOptimalLinear(p)
    \eqdef \arg\max_{s\in [0,1]} \inf_{c>0} \underline{F}(c,p,s).\label{eq:mfol_definition}
  \end{equation}
\end{enumerate}
As we will see shortly, the capacity-agnostic additive-gap optimal linear policy coincides with the fixed-fraction policy (Theorem~\ref{th:ag_minimax}), whereas the capacity-agnostic multiplicative-factor optimal linear policy is a new kind of capacity-agnostic linear policy.
The next equation gives a very precise approximation of $\camfOptimalLinear(p)$.
Let
\begin{equation}
  % eh4-8
  \camfOptimalLinearHat(p)
  \eqdef \min\left\{\frac{p}{2} \log(1+\tilde{s}(p)) + \left(1 - \frac{p}{2}\right) \tilde{s}(p), 1\right\},\label{eq:mfol_approximation}
\end{equation}
where
\(
\tilde{s}(p)
\eqdef (a^*)^{0.05} \log (1+(a^*)^{0.95}p)
\)
and $a^* \approx 2.2847$.
Numerical results show that $|\camfOptimalLinearHat(p)-\camfOptimalLinear(p)|<0.0015$ for all $p\in (0,1)$.

\subsection{From Quasi-Static Fading to Block Fading}\label{subsec:from_quasi_static_to_block_fading}

From a practical standpoint, the quasi-static-fading assumption in \eqref{eq:awgn} may be overly idealized.
A natural question is whether the optimal linear policies derived under this assumption remain effective under more general fading models.
To this end, consider a block-fading setting in which the channel gain (known at both the transmitter and receiver) remains constant within each time slot and varies independently and identically across slots according to a certain distribution.
In this setting, the reward in time slot $t$ is $r(G_t,\Gamma_t)$, where $G_t$ and $\Gamma_t$ denote the consumed energy and the random channel SNR coefficient in time slot $t$, respectively.
Analogous to Remark~\ref{snr_normalization_remark}, we assume $\expect\Gamma_t=1$ without loss of generality.

For $G_t\gg 1$ (the high-SNR regime),
\begin{equation}
  \expect r(G_t, \Gamma_t)
  \approx \expect \left(\frac{1}{2} \log (\Gamma_t + \Gamma_t G_t)\right)
  = r(G_t,1) + \frac{1}{2} \expect \log \Gamma_t,
\end{equation}
so $\expect r(G_t, \Gamma_t)$ is nearly $r(G_t, 1)$ plus a non-positive constant determined solely by the distribution of $\Gamma_t$.
Consequently, in this regime, the optimal linear policies derived for the quasi-static fading model are expected to remain near optimal under block fading.

However, in the low-SNR regime, the performance of linear battery-level-based policies can deteriorate.
On the one hand, the quasi-static surrogate $\expect r(G_t,\Gamma_t)\approx r(G_t,1) + \frac{1}{2} \expect \log \Gamma_t$ may be inaccurate, so linear policies optimized for the quasi-static fading model can suffer a noticeable throughput loss.
On the other hand, and more importantly, adapting the transmit power to (instantaneous) channel-power-gain information can yield substantially larger throughput gains in this regime.
Since linear policies do not exploit $\Gamma_t$, their gap to the optimal online policy can therefore be more pronounced at low SNR.

In this case, we can use dynamic programming to design improved policies, building on the optimal linear policies derived above.
Specifically, \emph{we approximate the future (next-slot) decision rule by a linear policy and, via a bootstrapping step, obtain an improved current-slot policy that explicitly exploits the channel-power-gain information}.
This idea has been realized in~\cite{wu2026clipped}.
As a result, we obtain the so-called \emph{robust clipped affine (RCA) policy}:
\begin{equation}
  \robustClippedAffinePolicy[p,q,\hat{\gamma}](b,\gamma)
  \eqdef \min\left\{\max\left\{\frac{qb-(1-p)/\gamma+1/\hat{\gamma}}{1-p+q},0\right\},b\right\}.\label{eq:rca_definition}
\end{equation}
Here, $q$ and $\hat{\gamma}$ denote the \emph{effectively equivalent linear-policy slope} and the \emph{effectively equivalent channel SNR coefficient} (for the next time slot), respectively.
Setting $\hat{\gamma}=1$ and taking $q\in\{\maximinOptimalLinear(c,p),\,\caagOptimalLinear(p),\,\camfOptimalLinear(p)\}$ yields three illustrative parameter choices of the robust clipped affine policy, which are expected to perform well across a wide range of SNR regimes.
For convenience, we denote the resulting policies by $\rcaOL[p]$, $\rcaAGOL[p]$, and $\rcaMFOL[p]$, respectively.
To some extent, this work paves the way for developing more advanced closed-form policies based on good linear policies.

\subsection{Simulation Results}\label{subsec:simulation_results}

In this subsection, we evaluate via simulations the performance of the three optimal linear policies $\maximinOptimalLinear$, $\caagOptimalLinear$, and $\camfOptimalLinear$, as well as their RCA counterparts $\rcaOL$, $\rcaAGOL$, and $\rcaMFOL$.
For comparison, we also include the maximin optimal policy $\maximinOptimalPolicy$~\cite[Thm.~1]{yang2020maximin} and the Lyapunov-optimization-based policy proposed in~\cite[Alg.~1]{amirnavaei2016online} (with $E_{\min}=0$, $E_{\max}=c$, $\Delta t P_{\max} = E_{\text{c,max}} = 0.2c$, and $\eta=0.01$), denoted by $\lyapunovPolicy$ and referred to as the Lyapunov policy.

To facilitate comparison across settings, we report the multiplicative factor $F_Q(c,\sigma)$ (Eq.~\eqref{eq:multiplicative_factor}) as the performance metric; equivalently, we also report the corresponding percentage performance loss, defined as $100(1-F_Q(c,\sigma))$.
For each setting, we first compute an optimal benchmark policy via policy iteration and then evaluate this policy by running it in the simulation to obtain the benchmark throughput $\mathcal{T}_c^*(Q)$.
We also follow the performance-evaluation framework in~\cite[Sec.~2.2.2]{yang2025power}, which is based on the following concepts:
\begin{itemize}
  \item \emph{Nominal mean-to-capacity ratio} (NMCR):
  \begin{equation}
    \text{NMCR}
    \eqdef \frac{\mu}{c},\label{eq:nmcr}
  \end{equation}
  where $\mu$ is the mean defined by~\eqref{eq:mean}.
  This ratio is easier to compute and use in practice than the MCR (Eq.~\eqref{eq:mcr}).
  However, energy-arrival distributions with the same NMCR may have different MCRs.
  This is illustrated by Table~\ref{tab:mcr}, which compares the MCRs of the three distribution families used in the simulation.
  For these families, their parameters, and consequently the MCR, are uniquely determined by the NMCR.
  \item \emph{Nominal signal-to-noise ratio} (NSNR) in decibels (dB):
  \begin{align}
    \text{NSNR}
    &\eqdef 10\log_{10} (\expect \Gamma_t \bar{\mu})
    = 10\log_{10} \bar{\mu}, \label{eq:nsnr}
  \end{align}
  where $\Gamma_t$ denotes the channel SNR coefficient and $\bar{\mu}$ is the clipped mean defined by~\eqref{eq:clipped_mean}.
  Given the NSNR and MCR, the battery capacity $c$ can be computed by
  \begin{equation}
    c
    = \frac{\bar{\mu}}{\text{MCR}}
    = \frac{10^{\text{NSNR/10}}}{\text{MCR}}.
  \end{equation}
\end{itemize}

\begin{table}
  \centering
  \renewcommand{\arraystretch}{1.2}
  \caption{MCRs of Bernoulli, Exponential, and Uniform Distributions \cite[Table~2.2]{yang2025power}}\label{tab:mcr}
  \begin{tabular}{ccccc}
    \hline
    Distribution & MCR for NMCR $\tilde{p}$ & $\tilde{p}=0.1$ & $0.5$ & $0.9$ \\
    \hline
    Bernoulli    & $\tilde{p}$              & $0.1$           & $0.5$ & $0.9$ \\
    Exponential & $\tilde{p}(1 - \en^{-1/\tilde{p}})$
    & $0.1000$ & $0.4323$ & $0.6037$ \\[0.5ex]
    Uniform & \(
    \begin{cases}
      \tilde{p},               & \(\tilde{p} \in [0,\frac{1}{2}]\) \\
    1-\dfrac{1}{4\tilde{p}}, & \(\tilde{p}>\frac{1}{2}\)
  \end{cases}\) & $0.1$ & $0.5$ & $0.7222$ \\[4ex]
  \hline
\end{tabular}
\end{table}

\begin{figure*}
  \centering
  \includegraphics{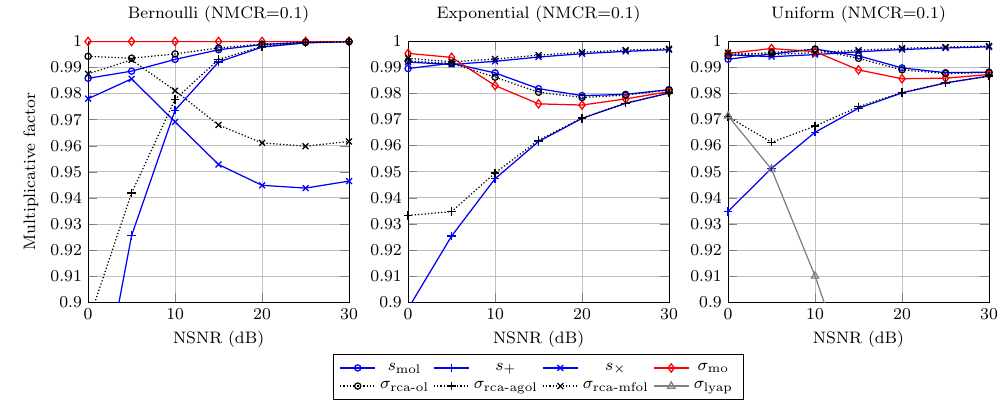}
  \caption{Multiplicative-factor performance comparison for quasi-static fading and three energy-arrival distribution families with $\text{NMCR}=0.1$.}\label{fig:awgn-bern-mf}
\end{figure*}

\begin{figure*}
  \centering
  \includegraphics{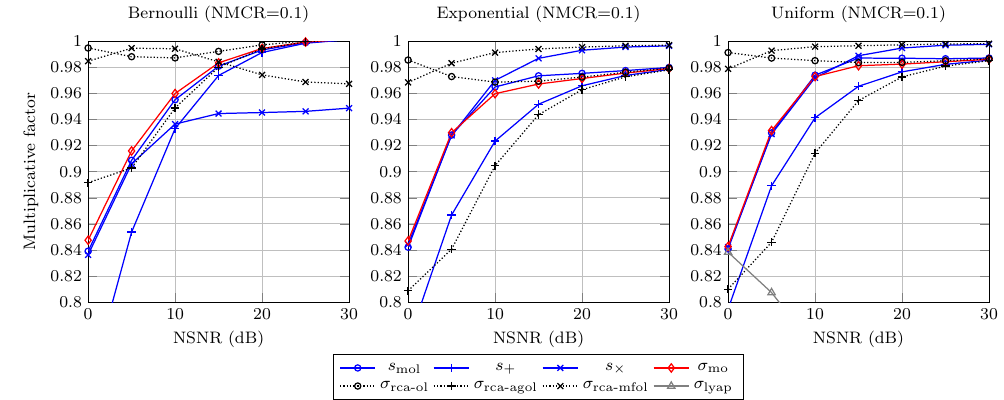}
  \caption{Multiplicative-factor performance comparison for Rayleigh block fading and three energy-arrival distribution families with $\text{NMCR}=0.1$.}\label{fig:rayleigh-bern-mf}
\end{figure*}

The simulation is conducted under two fading models: quasi-static fading and Rayleigh block fading.
For each fading model, we consider three families of energy-arrival distributions: Bernoulli, exponential, and uniform distributions.
For each combination of fading model and energy-arrival distribution family, we set $\text{NMCR} \in \{0.1, 0.5, 0.9\}$ and sweep the NSNR from $0$~dB to $30$~dB in $5$~dB increments.
Each setting is simulated for \(10^3\) episodes of \(10^4\) slots each, with a uniformly random initial battery level.
\begin{table}
  \centering
  \renewcommand{\arraystretch}{1.2}
  \caption{Percentage Performance Loss Relative to the Optimal Policies}\label{tab:performance_loss}
  \begin{tabular}{ccccc}
    \toprule
    & \multicolumn{2}{c}{Quasi-Static Fading} & \multicolumn{2}{c}{Rayleigh Block Fading} \\
    \cline{2-3}\cline{4-5}
    Policy                  & Average   & Maximum   & Average   & Maximum   \\
    \midrule
    $\maximinOptimalLinear$ & $0.92\%$  & $2.09\%$  & $3.01\%$  & $16.10\%$ \\
    $\caagOptimalLinear$    & $3.72\%$  & $16.29\%$ & $5.56\%$  & $26.84\%$ \\
    $\camfOptimalLinear$    & $1.28\%$  & $5.62\%$  & $3.19\%$  & $16.37\%$ \\
    $\maximinOptimalPolicy$ & $0.77\%$  & $2.45\%$  & $2.92\%$  & $15.74\%$ \\
    \hline
    $\rcaOL$                & $0.87\%$  & $2.23\%$  & $1.41\%$  & $3.65\%$  \\
    $\rcaAGOL$              & $1.85\%$  & $10.77\%$ & $3.63\%$  & $19.12\%$ \\
    $\rcaMFOL$              & $0.78\%$  & $4.02\%$  & $1.07\%$  & $3.27\%$  \\
    $\lyapunovPolicy$       & $14.18\%$ & $26.81\%$ & $27.32\%$ & $37.42\%$ \\
    \bottomrule
  \end{tabular}
\end{table}
The simulation results are summarized in Table~\ref{tab:performance_loss}, which reports the average and maximum percentage performance loss relative to the optimal policies over all settings.
Figures~\ref{fig:awgn-bern-mf} and~\ref{fig:rayleigh-bern-mf} illustrate the multiplicative-factor performance comparison under quasi-static fading and Rayleigh block fading, respectively, for $\text{NMCR}=0.1$.

In the quasi-static fading setting, the maximin optimal linear policy $\maximinOptimalLinear$ performs comparably to the maximin optimal policy $\maximinOptimalPolicy$, with an average performance loss below $1\%$ and a maximum loss of about $2\%$.
The capacity-agnostic multiplicative-factor optimal linear policy $\camfOptimalLinear$ performs slightly worse than $\maximinOptimalLinear$, but remains significantly better than the capacity-agnostic additive-gap optimal linear policy $\caagOptimalLinear$ (i.e., the fixed-fraction policy).
The latter performs well in the high-NSNR regime but suffers from a noticeable performance loss in the low-NSNR regime.

In the Rayleigh block fading setting, the three optimal linear policies still maintain satisfactory performance at high NSNR, but their performance degrades significantly at low NSNR.
This is expected, as explained in Section~\ref{subsec:from_quasi_static_to_block_fading}.
By contrast, the RCA policies significantly improve the performance in this setting, especially in low NSNR regimes.
Among them, $\rcaOL$ and $\rcaMFOL$ achieve the best performance, with average performance losses below $1.5\%$ and maximum losses below $4\%$.

Overall, the maximin optimal linear and capacity-agnostic multiplicative-factor optimal linear policies, together with their RCA counterparts, consistently rank among the top-performing policies across the compared settings.

The Lyapunov policy $\lyapunovPolicy$ is evaluated only for the uniform energy-arrival distribution with $\text{NMCR}=0.1$, since its applicability relies on stringent constraints on the maximum charging and discharging energy per time slot.
Its performance is markedly worse than that of the other policies, especially under Rayleigh block fading.
This is because it is derived from a relaxed optimization problem that removes the battery constraint~\cite[P2]{amirnavaei2016online}; consequently, it tends to perform well only for energy supply with very small fluctuations.

\section{Performance Analysis}\label{sec:performance_analysis}

This section presents a theoretical performance analysis of the optimal linear policies, providing a rigorous foundation for the simulation results shown in the previous section.
It also investigates the optimality of the greedy policy for certain energy-arrival distribution families, including when only the battery capacity $c$ and the MCR $p$ are known.

\subsection{Saddle-Point Structures of Maximin Optimal Linear Policies}\label{subsec:saddle_point_structures}

In this subsection, we characterize the worst nominal additive gaps and the worst nominal multiplicative factors of the three optimal linear policies from Section~\ref{sec:worst_case_optimal_linear_policies}.
A key insight is that the nominal additive gap and the nominal multiplicative factor each induce a saddle-point structure over $(c,p,s)$.
In particular, the capacity-agnostic additive-gap-optimal and multiplicative-factor-optimal linear policies constitute corresponding saddle points in the extended and strict senses, respectively.
Due to the difficulty of the problem, some claims cannot be proved and are therefore supported by extensive numerical verification.
These claims are stated as conjectures and used as assumptions in the subsequent theorems.

First, we present three results on the quasiconvexity or quasiconcavity of the worst-case throughput, the nominal additive gap, and the nominal multiplicative factor for linear policies.

\begin{conjecture}\label{cj:throughput_quasi_concave_in_s}
  The worst-case throughput $\underline{\Gamma}(c,p,s)$ is strictly quasiconcave in $s$ for fixed $p\in (0,1)$ and $c>0$.
  \verificationref{Appendix~\ref{sec:numerical_verification}}
\end{conjecture}

\begin{proposition}\label{pr:ag_increasing_in_c}
  Fix $p\in (0,1)$.
  The nominal additive gap $\overline{G}(c,p,s)$ is strictly increasing in $c$ for fixed $s\in [0,1]$; the same holds when $s=\maximinOptimalLinear(c,p)$.
  Moreover, for fixed $s\in [0,1]$,
  \begin{equation}
    \lim_{c\to +\infty} \overline{G}(c,p,s)
    = G_p(s)
    \eqdef \frac{1}{2} \left(\log\frac{p}{s} - \frac{1-p}{p}\log(1-s)\right). \label{g_p}
  \end{equation}
  \proofref{Appendix~\ref{sec:proofs_of_saddle_point_structures}}
\end{proposition}

\begin{conjecture}\label{cj:mf_quasi_convex_in_c}
  For any $p\in (0,1)$ and $s\in (0,1]$, the nominal multiplicative factor $\underline{F}(c,p,s)$ is strictly quasiconvex in $c$.
  \verificationref{Appendix~\ref{sec:numerical_verification}}
\end{conjecture}

Next, we present the main results on the saddle-point structures of the optimal linear policies.

\begin{theorem}\label{th:ag_minimax}
  Fix $p\in (0,1)$.
  Then,
  \begin{equation}
    \min_{s\in [0,1]} \sup_{c>0} \overline{G}(c,p,s)
    = \sup_{c>0} \min_{s\in [0,1]} \overline{G}(c,p,s)
    = G_p(p),\label{ag_saddle_point}
  \end{equation}
  where $G_p(s)$ is defined by~\eqref{g_p}.
  Positive infinity is the unique ``maximizer'' of $\min_{s\in [0,1]} \overline{G}(c,p,s)$ in $c$, i.e.,
  \begin{equation}
    \lim_{c\to +\infty} \min_{s\in [0,1]} \overline{G}(c,p,s)
    = \sup_{c>0} \min_{s\in [0,1]} \overline{G}(c,p,s).
  \end{equation}
  The slope $\caagOptimalLinear(p)=p$ is the unique minimizer of $\sup_{c>0} \overline{G}(c,p,s)$ in $s$.
  Moreover,
  \begin{equation}
    \lim_{c\to +\infty} \maximinOptimalLinear(c,p) = p,
  \end{equation}
  which implies that $(+\infty, \caagOptimalLinear(p))$ is the unique saddle point in the extended sense.
  \proofref{Appendix~\ref{sec:proofs_of_saddle_point_structures}}
\end{theorem}

\begin{theorem}\label{th:mf_minimax}
  Fix $p\in (0,1)$.
  If Conjectures~\ref{cj:throughput_quasi_concave_in_s} and~\ref{cj:mf_quasi_convex_in_c} hold, then
  \begin{equation}
    F_p
    \eqdef \max_{s\in [0,1]} \inf_{c>0} \underline{F}(c,p,s)
    = \inf_{c>0} \max_{s\in [0,1]} \underline{F}(c,p,s).\label{eq:mf_saddle_point}
  \end{equation}
  Moreover,
  \begin{equation}
    c_{\times}(p)\eqdef \arg\inf_{c>0} \max_{s\in [0,1]} \underline{F}(c,p,s) \label{eq:c_times}
  \end{equation}
  exists (i.e., a unique finite minimizer exists), and $(c_{\times}(p), \camfOptimalLinear(p))$ is the unique saddle point and satisfies $\camfOptimalLinear(p)
  = \maximinOptimalLinear(c_{\times}(p),p)$.
\end{theorem}

Unlike the nominal additive gap in Theorem~\ref{th:ag_minimax}, the minimax value $F_p$ (Eq.~\eqref{eq:mf_saddle_point}) of the nominal multiplicative factor does not appear to admit a closed-form expression, making it difficult to characterize $\inf_{p\in(0,1)} F_p$ analytically.
The next conjecture, supported by extensive numerical evidence, offers a convenient route to characterizing $\inf_{p\in(0,1)} F_p$.

\begin{conjecture}\label{cj:worst_mf_at_zero}
  The infimum of $F_p$ over $p\in(0,1)$ is attained in the limit as $p\to 0$.
  As $p\to 0$, the limits $\lim_{p\to 0} pc_{\times}(p)$ and $\lim_{p\to 0} \camfOptimalLinear(p)/p$ exist and are strictly positive, where $c_\times(p)$ is defined by~\eqref{eq:c_times}.
  \verificationref{Appendix~\ref{sec:numerical_verification}}
\end{conjecture}

We thus turn to characterizing the asymptotic behavior of $\underline{\Gamma}(b/p, p, ap)$ as $p\to 0$ for fixed $a>0$ and $b>0$, which is the key to characterizing $\lim_{p\to 0} F_p$.

\begin{lemma}\label{le:cs_asymptotic}
  Let $a>0$, $b>0$, and $0<p<1/\max\{a,1\}$.
  Then
  \begin{equation}
    - p\max\{a,1\}r(ab)
    \le \underline{\Gamma}(b/p, p, ap) - \Gamma_0(a,b)
    \le p\left(r(ab) + \frac{a}{2}\min\{ab, b, 1\}\right),
  \end{equation}
  where
  \begin{equation}
    \Gamma_0(a,b)
    \eqdef \int_0^{+\infty} \en^{-x} r(abe^{-ax}) \diff x
    = \int_0^1 r(abx^a) \diff x.\label{eq:gamma_0}
  \end{equation}
  \proofref{Appendix~\ref{sec:proofs_of_saddle_point_structures}}
\end{lemma}

We are now ready to characterize $\inf_{p\in(0,1)} F_p$.

\begin{theorem}\label{th:worst_mf}
  If Conjecture~\ref{cj:worst_mf_at_zero} holds, then
  \begin{gather}
    \inf_{p\in (0,1)} F_p
    = \sup_{a>0} \inf_{b>0} \frac{\Gamma_0(a,b)}{r(b)}
    = \frac{\Gamma_0(a^*,b^*)}{r(b^*)}
    \approx 0.6530,\\
    \lim_{p\to 0} \frac{\camfOptimalLinear(p)}{p}
    = a^*, \label{eq:a_star_limit}\\
    \lim_{p\to 0} pc_{\times}(p)
    = b^*,
  \end{gather}
  where $\Gamma_0(a,b)$ is defined by~\eqref{eq:gamma_0}, $c_\times(p)$ is defined by~\eqref{eq:c_times}, and
  \begin{gather}
    a^*
    \eqdef \arg \sup_{a>0} \inf_{b>0} \frac{\Gamma_0(a,b)}{r(b)}
    \approx 2.2847, \label{eq:a_star}\\
    b^*
    \eqdef \arg \inf_{b>0} \frac{\Gamma_0(a^*,b)}{r(b)}
    \approx 1.7938. \label{eq:b_star}
  \end{gather}
\end{theorem}

With the above characterization in place, we can now derive the worst nominal additive gaps and the worst nominal multiplicative factors of the three optimal linear policies.

\begin{theorem}\label{th:mol_worst_performance}
  The worst nominal additive gap of the maximin optimal linear policy $\maximinOptimalLinear$ is
  \[
    \sup_{c>0, p\in (0,1)} \overline{G}(c,p,\maximinOptimalLinear(c,p))
    = \frac{1}{2}.
  \]
  If Conjectures~\ref{cj:throughput_quasi_concave_in_s}--\ref{cj:worst_mf_at_zero} holds, then the worst nominal multiplicative factor of $\maximinOptimalLinear$ is
  \begin{equation}
    \inf_{c>0, p\in (0,1)} \underline{F}(c,p,\maximinOptimalLinear(p))
    = \frac{\Gamma_0(a^*,b^*)}{r(b^*)}
    \approx 0.6530,
  \end{equation}
  where $\Gamma_0$, $a^*$, and $b^*$ are defined by~\eqref{eq:gamma_0}, \eqref{eq:a_star} and~\eqref{eq:b_star}, respectively.
\end{theorem}

\begin{sketch}
  By the minimax identities in Theorems~\ref{th:ag_minimax} and~\ref{th:mf_minimax}, the worst nominal additive gap and multiplicative factor are achieved by the saddle-point policies $\caagOptimalLinear$ and $\camfOptimalLinear$, respectively.
  Thus, the worst values for $\maximinOptimalLinear$ match those of $\caagOptimalLinear$ (additive gap) and $\camfOptimalLinear$ (multiplicative factor).
  The claim then follows from~\eqref{eq:agol_worst_ag} and~\eqref{eq:mfol_worst_mf}.
\end{sketch}

\begin{theorem}[{\cite[Props.~3 and~4]{shaviv2016universally} and~\cite[Thms.~29 and~30]{yang2020.6maximin}}]\label{th:agol_worst_performance}
  The worst nominal additive gap of the capacity-agnostic additive-gap optimal linear policy $\caagOptimalLinear$ is
  \begin{equation}
    \sup_{c>0, p\in (0,1)} \overline{G}(c,p,\caagOptimalLinear(p))
    = \frac{1}{2}.\label{eq:agol_worst_ag}
  \end{equation}
  The worst nominal multiplicative factor of $\caagOptimalLinear$ is
  \begin{equation}
    \inf_{c>0, p\in (0,1)} \underline{F}(c,p,\caagOptimalLinear(p))
    = \frac{1}{2}.\label{eq:agol_worst_mf}
  \end{equation}
\end{theorem}

\begin{sketch}
  Apply Theorem~\ref{th:ag_minimax} and evaluate $\sup_{p\in (0,1)} G_p(p) = \lim_{p\to 0} G_p(p)$ to derive~\eqref{eq:agol_worst_ag}.
  Use Jensen's inequality to obtain $\underline{F}(c,p,\caagOptimalLinear(p)) \ge \frac{1}{2}$ for all $c>0$ and $p\in (0,1)$, and show that this lower bound is achievable, leading to~\eqref{eq:agol_worst_mf}.
\end{sketch}

While Eqs.~\eqref{eq:agol_worst_ag} and~\eqref{eq:agol_worst_mf} are well-established results, no single paper provides a complete proof.
The reader is referred to~\cite[Sec.~3.4.2]{yang2025power} for a detailed and consolidated proof.

\begin{theorem}\label{th:mfol_worst_performance}
  % eh4-18
  The worst nominal additive gap of the capacity-agnostic multiplicative-factor optimal linear policy $\camfOptimalLinear$ is
  \begin{equation}
    \sup_{c>0, p\in (0,1)} \overline{G}(c,p,\camfOptimalLinear(p))
    = \sup_{p\in (0,1)} G_p(\camfOptimalLinear(p))
    \approx 0.7292.\label{eq:mfol_worst_ag}
  \end{equation}
  If $G_p(\camfOptimalLinear(p))$ attains its supremum in the limit as $p\to 0$, then
  \begin{equation}
    \sup_{p\in (0,1)} G_p(\camfOptimalLinear(p))
    = \limsup_{p\to 0} \frac{1}{2}\left(\frac{\camfOptimalLinear(p)}{p} - \log\frac{\camfOptimalLinear(p)}{p}\right),\label{eq:mfol_worst_ag_limit}
  \end{equation}
  which agrees with the numerical result in~\eqref{eq:mfol_worst_ag} under the assumption that~\eqref{eq:a_star_limit} holds.
  If Conjecture~\ref{cj:worst_mf_at_zero} holds, then the worst nominal multiplicative factor of $\camfOptimalLinear$ is
  \begin{equation}
    \inf_{c>0, p\in (0,1)} \underline{F}(c,p,\camfOptimalLinear(p))
    = \frac{\Gamma_0(a^*,b^*)}{r(b^*)}
    \approx 0.6530,\label{eq:mfol_worst_mf}
  \end{equation}
  where $\Gamma_0$, $a^*$, and $b^*$ are defined by~\eqref{eq:gamma_0}, \eqref{eq:a_star} and~\eqref{eq:b_star}, respectively.
\end{theorem}

\begin{sketch}
  Use Proposition~\ref{pr:ag_increasing_in_c} to establish~\eqref{eq:mfol_worst_ag}, and derive~\eqref{eq:mfol_worst_ag_limit} with the given assumption.
  Apply Theorems~\ref{th:mf_minimax} and~\ref{th:worst_mf} to conclude~\eqref{eq:mfol_worst_mf}.
\end{sketch}

\subsection{Optimality of Greedy Policy for Certain Families of Energy-Arrival Distributions}\label{subsec:greedy_policy_optimality}

In~\cite{wang2021optimality}, the greedy policy is shown to maximize the throughput in the small battery capacity regime.
Specifically, for a reward function $r(x)$, the greedy policy is optimal if and only if
\begin{equation}
  c
  \le c^*(Q)
  \eqdef \max\left\{c\ge 0: r'(c)\ge \int_{[0,c)} r'(x) \diff Q\right\},
  \label{greedy_threshold}
\end{equation}
where $Q$ denotes the marginal distribution of the i.i.d.\ process of energy arrivals, and $c^*(Q)$ is the \emph{battery-capacity threshold for greedy optimality} (hereafter the \emph{greedy threshold}) for $Q$.
For the reward function \eqref{eq:awgn}, the threshold is given by
\begin{equation} \label{grdy:thr_cap}
  c^*(Q)
  = \max\left\{c\ge 0: \frac{1}{1+c}\geq \int_{[0,c)}\frac{1}{1+x} \diff Q  \right\}.
\end{equation}
As an example, we have $c^*(\bernoulli_{1/(1-p),p}) = p/(1-p)$ (cf.\ Proposition~\ref{pr:mol.special}).

However, as the exact distribution of the energy-arrival process is not always in hand, our aim in the sequel is to find the tightest semi-universal bounds on the greedy threshold for certain families of energy-arrival distributions.
Specifically, we are interested in three cases, from general to special:

\begin{enumerate}
  \item distribution $Q$ with the possible-value interval $[\underline{x},\overline{x}]$ (satisfying $Q([\underline{x},\overline{x}]) = 1$) and the mean $\mu$ ($\underline{x}\le \mu\le \overline{x}$);

  \item clipped distribution $Q$ with the possible-value interval $[\underline{x}, c]$ and the clipped mean $\bar{\mu}$ ($\underline{x}\le \bar{\mu}\le c$);

  \item clipped distribution $Q$ with the possible-value interval $[0, c]$ and the MCR $p\in (0,1)$.
\end{enumerate}

\subsubsection{Greedy Threshold Bounds for Given Possible-Value Interval and Mean}
\label{greedy_policy.1}

We determine the tightest lower and upper bounds on the greedy threshold given the possible-value interval $[\underline{x}, \overline{x}]$ and the mean $\mu$.

Let
\begin{gather}
  \underline{c}(\underline{x},\overline{x},\mu)
  \eqdef \inf_{Q\in Q_{\underline{x},\overline{x},\mu}} c^*(Q),
  \quad
  \overline{c}(\underline{x},\overline{x},\mu)
  \eqdef \sup_{Q\in Q_{\underline{x},\overline{x},\mu}} c^*(Q),
\end{gather}
where
\(
\mathcal{Q}_{\underline{x},\overline{x},\mu}
\eqdef \{Q: Q([\underline{x},\overline{x}])=1, \expect_{X\sim Q}X = \mu\}.
\)
In order to find $\underline{c}(\underline{x},\overline{x},\mu)$ and $\overline{c}(\underline{x},\overline{x},\mu)$, one must maximize (minimize) the integral in \eqref{greedy_threshold}.
Thus, we need to find the values of
\[
  \overline{f}(c,\underline{x},\overline{x},\mu)
  \eqdef \sup_{Q \in \mathcal{Q}_{\underline{x},\overline{x},\mu}} \int_{[0,c)} r'(x) \diff Q
\]
and
\[
  \underline{f}(c,\underline{x},\overline{x},\mu)
  \eqdef \inf_{Q \in \mathcal{Q}_{\underline{x},\overline{x},\mu}} \int_{[0,c)} r'(x) \diff Q.
\]
The relation between $\underline{c}(\underline{x},\overline{x},\mu)$ (resp., $\overline{c}(\underline{x},\overline{x},\mu)$) and $\overline{f}(c,\underline{x},\overline{x},\mu)$ (resp., $\underline{f}(c,\underline{x},\overline{x},\mu)$) is established by the next lemma.

\begin{lemma}\label{f_bound_to_c_bound}
  Let $r$ be a non-decreasing, continuously differentiable, and strictly concave function on $[0,+\infty)$.%
  \footnote{%
    In order to apply \eqref{greedy_threshold}, we require that the reward function $r$ satisfy \cite[Assumptions~1 and 2]{wang2021optimality} for all $Q\in \mathcal{Q}_{\underline{x},\overline{x},\mu}$, so $r$ must be non-decreasing, continuously differentiable, and in particular, strictly concave (at least on $[\underline{x},\overline{x}]$).
  }
  Then,
  \begin{gather}
    \underline{c}(\underline{x},\overline{x},\mu)
    = \underline{c}'(\underline{x},\overline{x},\mu)
    \eqdef \sup \left\{c\ge 0: r'(c)\ge \overline{f}(c,\underline{x},\overline{x},\mu)\right\},\\
    \overline{c}(\underline{x},\overline{x},\mu)
    = \overline{c}'(\underline{x},\overline{x},\mu)
    \eqdef \sup \left\{c\ge 0: r'(c)\ge \underline{f}(c,\underline{x},\overline{x},\mu)\right\}.
  \end{gather}
  \proofref{Appendix~\ref{sec:proofs_of_greedy_policy_optimality}}
\end{lemma}

In the rest of this subsection, we will focus on the case of reward function \eqref{eq:awgn}.
In this case, the exact values of $\underline{f}(c,\underline{x},\overline{x},\mu)$ and $\overline{f}(c,\underline{x},\overline{x},\mu)$ are determined by the next lemma.

\begin{lemma}\label{f_bounds}
  \begin{equation}
    2\underline{f}(c,\underline{x},\overline{x},\mu)
    = \begin{cases}
        0 &$c\in [0,\mu]$,\\
        \frac{c-\mu}{(1+\underline{x})(c-\underline{x})} &$c\in [0, \iota(\underline{x})) \cap (\mu,\overline{x}]$,\\
        \frac{4(c-\mu)}{(c+1)^2} &$c\in [\iota(\underline{x}),\iota(\mu)) \cap (\mu,\overline{x}]$,\\
        \frac{1}{1+\mu} &$c\in [\iota(\mu),\overline{x}]\cup (\overline{x},+\infty)$,
    \end{cases}\label{f_bounds.lower}
  \end{equation}
  \begin{equation}
    2\overline{f}(c,\underline{x},\overline{x},\mu)
    = \begin{cases}
        0 &$c\in [0,\underline{x}]$,\\
        \frac{\overline{x}-\mu}{(\overline{x}-\underline{x})(1+\underline{x})} &$c\in [0,\tau]\cap (\underline{x},\overline{x}]$,\\
        \frac{\overline{x}-\mu}{(\overline{x}-c)(1+c)} &$c\in (\tau,+\infty)\cap (\underline{x},\mu]$,\\
        \frac{1+\underline{x}+c-\mu}{(1+\underline{x})(1+c)} &$c\in (\tau,+\infty)\cap (\mu,\overline{x}]$,\\
        \frac{1+\underline{x}+\overline{x}-\mu}{(1+\underline{x})(1+\overline{x})} &$c\in (\overline{x},+\infty)$,\\
    \end{cases}\label{f_bounds.upper}
  \end{equation}
  where $\iota(x)\eqdef 2x+1$ and $\tau\eqdef \overline{x}-\underline{x}-1$.
  \proofref{Appendix~\ref{sec:proofs_of_greedy_policy_optimality}}
\end{lemma}

Based on Lemma~\ref{f_bounds}, we obtain $\underline{c}(\underline{x},\overline{x},\mu)$ and $\overline{c}(\underline{x},\overline{x},\mu)$.

\begin{theorem}\label{thm:grdy:lower_c}
  \begin{equation}
    \underline{c}(\underline{x},\overline{x},\mu)
    = \begin{cases}
        \underline{c}_1 &$\mu<\tau$,\\
        \mu &$\mu\ge\tau$,
    \end{cases}
  \end{equation}
  where $\tau\eqdef \overline{x}-\underline{x}-1$ and
  \begin{equation}
    \underline{c}_1
    \eqdef \frac{(\overline{x}-\underline{x})(1+\underline{x})}{\overline{x}-\mu}-1.
  \end{equation}
  \proofref{Appendix~\ref{sec:proofs_of_greedy_policy_optimality}}
\end{theorem}

\begin{corollary}\label{lower_c_eq_upper_x}
  $\underline{c}(\underline{x},\overline{x},\mu) = \overline{x}$ if and only if $\mu=\overline{x}$.
\end{corollary}

\begin{theorem}\label{thm:grdy:upper_c}
  \begin{equation}
    \overline{c}(\underline{x},\overline{x},\mu)
    = \begin{cases}
        \min\{\overline{c}_1,\overline{x}\} &$\mu<\frac{3}{2}\underline{x}+\frac{1}{2}$,\\
        \min\{\overline{c}_2,\overline{x}\} &$\mu\ge\frac{3}{2}\underline{x}+\frac{1}{2}$,
    \end{cases}
  \end{equation}
  where
  \begin{gather}
    \overline{c}_1
    \eqdef \frac{\underline{x}+\mu+\sqrt{(\underline{x}+\mu)^2-4(\underline{x}^2+\underline{x}-\mu)}}{2},\\
    \overline{c}_2 \eqdef \frac{4\mu+1}{3}.
  \end{gather}
  \proofref{Appendix~\ref{sec:proofs_of_greedy_policy_optimality}}
\end{theorem}

\begin{corollary}\label{upper_c_eq_upper_x}
  $\overline{c}(\underline{x},\overline{x},\mu) = \overline{x}$ if and only if
  \begin{equation}
    \overline{x}
    \le \begin{cases}
          \overline{c}_1 &$\mu<\frac{3}{2}\underline{x}+\frac{1}{2}$,\\
          \overline{c}_2 &$\mu\ge\frac{3}{2}\underline{x}+\frac{1}{2}$.
    \end{cases}
  \end{equation}
\end{corollary}

\begin{remark}
  The bounds given by Theorems~\ref{thm:grdy:lower_c} and \ref{thm:grdy:upper_c} coincide with the bounds in \cite[Props.~4 and 5]{wang2021optimality}, and hence the tightness of the former implies the tightness of the latter.
  This observation can be easily understood by the following trick.
  Let $Q$ be a distribution attaining $\underline{c}(\underline{x},\overline{x},\mu)$ or $\overline{c}(\underline{x},\overline{x},\mu)$.
  By definition, we only have $Q([\underline{x}, \overline{x}]) = 1$, and in general, the essential infimum and supremum of a random variable with distribution $Q$ may be strictly larger than $\underline{x}$ and strictly less than $\overline{x}$, respectively.
  Consider a random variable $X_t$ with distribution
  \[
    Q_{t}
    \eqdef (1-t)Q + t \left( \frac{\overline{x}-\mu}{\overline{x}-\underline{x}} \delta_{\underline{x}} + \frac{\mu-\underline{x}}{\overline{x}-\underline{x}} \delta_{\overline{x}} \right)
  \]
  where $t\in (0,1]$.
  Then, the essential minimum and maximum of $X_t$ are $\underline{x}$ and $\overline{x}$, respectively.
  Taking $t=1/n$, we obtain a sequence $\{Q_{1/n}\}_{n=1}^\infty$ of distributions approaching the bounds in \cite[Props.~4 and 5]{wang2021optimality}.
\end{remark}

\subsubsection{Greedy Threshold Bounds for Given Least Possible Value and Clipped Mean}
\label{greedy_policy.2}

We determine the tightest lower and upper bounds on the greedy threshold given the least possible value $\underline{x}$ and the clipped mean $\bar{\mu}$.

Let
\begin{gather}
  \underline{c}'(\underline{x},\bar{\mu})
  \eqdef \inf \{c\ge\bar{\mu}: c^*(Q)\ge c\ \text{for all $Q\in \mathcal{Q}_{\underline{x},c,\bar{\mu}}$}\},\\
  \overline{c}'(\underline{x},\bar{\mu})
  \eqdef \sup \{c\ge\bar{\mu}: c^*(Q)\ge c\ \text{for some $Q\in \mathcal{Q}_{\underline{x},c,\bar{\mu}}$}\}.
\end{gather}
By Corollaries~\ref{lower_c_eq_upper_x} and \ref{upper_c_eq_upper_x} as well as the proof of Lemma~\ref{f_bounds}, it is easy to determine the values of $\underline{c}'(\underline{x},\bar{\mu})$ and $\overline{c}'(\underline{x},\bar{\mu})$.

\begin{theorem}\label{lower_upper_c_for_clipped}
  \begin{gather}
    \underline{c}'(\underline{x},\bar{\mu})
    = \bar{\mu},\\
    \overline{c}'(\underline{x},\bar{\mu})
    = \begin{cases}
        \overline{c}_1 &$\bar{\mu}<\frac{3}{2}\underline{x}+\frac{1}{2}$,\\
        \overline{c}_2 &$\bar{\mu}\ge\frac{3}{2}\underline{x}+\frac{1}{2}$,
    \end{cases}
  \end{gather}
  where $\overline{c}'(\underline{x},\bar{\mu})$ is attained by
  \begin{equation}
    \overline{Q}_1
    \eqdef \begin{cases}
             \frac{c-\bar{\mu}}{c-\underline{x}} \delta_{\underline{x}} + \frac{\bar{\mu}-\underline{x}}{c-\underline{x}} \delta_c &$\bar{\mu}<\frac{3}{2}\underline{x}+\frac{1}{2}$,\\
             \frac{2(c-\bar{\mu})}{c+1}\delta_{(c-1)/2} + \frac{2\bar{\mu}-c+1}{c+1}\delta_c &$\bar{\mu}\ge\frac{3}{2}\underline{x}+\frac{1}{2}$.
    \end{cases}
  \end{equation}
\end{theorem}

\subsubsection{Greedy Threshold Upper Bound for Given MCR}
\label{greedy_policy.3}

We determine the tightest upper bound on the greedy threshold given the MCR $p$.

\begin{theorem}\label{upper_c_for_mcr}
  Let $\overline{c}''(p)\eqdef \sup \{c>0: c\le \overline{c}'(0,pc)\}$.
  Then
  \begin{equation}
    \overline{c}''(p)
    = \begin{cases}
        \frac{p}{1-p} &$p\in (0,\frac{1}{2})$,\\
        \frac{1}{3-4p} &$p\in [\frac{1}{2},\frac{3}{4})$,\\
        +\infty &$p\in [\frac{3}{4},1)$,
    \end{cases}
  \end{equation}
  which is attained by
  \begin{equation}
    \overline{Q}_2
    \eqdef \begin{cases}
             (1-p) \delta_0 + p \delta_{p/(1-p)} &$p\in (0,\frac{1}{2})$,\\
        \frac{1}{2}\delta_{(2p-1)/(3-4p)} + \frac{1}{2}\delta_{1/(3-4p)} &$p\in [\frac{1}{2},\frac{3}{4})$,
    \end{cases}
  \end{equation}
  and $\{\overline{Q}_3^{(n)}\}_{n=1}^\infty$ (for the last case) with
  \begin{equation}
    \overline{Q}_3^{(n)}
    \eqdef \frac{2n(1-p)}{n+1}\delta_{(n-1)/2} + \frac{2pn-n+1}{n+1}\delta_n.
  \end{equation}
  \proofref{Appendix~\ref{sec:proofs_of_greedy_policy_optimality}}
\end{theorem}

\section{Conclusion}
\label{sec:conclusion}

We have systematically investigated linear power control policies for energy harvesting communications. Our formulations require a minimal amount of information regarding the energy-arrival process, and consequently can capture various universality aspects of linear policies. The analysis of such formulations is feasible largely due to certain extremal properties of the Bernoulli energy-arrival process and its variants. As shown in \cite{yang2020maximin}, to some extent, these extremal properties continue to be preserved even when a broader class of policies (not necessarily linear) are adopted. So it might be possible to expand the scope of our work by going beyond linear policies, which will enable a meaningful discussion of
complexity vs. performance in the context of online power control.

\appendices

\section{Proofs of Results in Section~\ref{subsec:saddle_point_structures}}\label{sec:proofs_of_saddle_point_structures}

\begin{proofof}{Proposition~\ref{pr:ag_increasing_in_c}}
  We first compute the partial derivative of $\overline{G}(c,p,s)$ with respect to $c$.
  For fixed $s$, define
  \[
    f(c, p, s)
    \eqdef \frac{\partial \overline{G}(c,p,s)}{\partial c}
    = pr'(pc)-\sum_{i=0}^\infty p(1-p)^i r'(cs(1-s)^i) s(1-s)^i.
  \]
  For $s = \maximinOptimalLinear(c,p)$, we have
  \[
    \frac{\partial \overline{G}(c,p,s)}{\partial c}
    = f(c,p,s) - \frac{\partial \underline{\Gamma}(c,p,s)}{\partial s}\Bigg|_{s=\maximinOptimalLinear(c,p)} \cdot \frac{\partial \maximinOptimalLinear(c,p)}{\partial c}
    \eqvar{(a)} f(c,p,s),
  \]
  where (a) is justified as follows.
  If $c\in [0,p/(1-p)]$, then $\maximinOptimalLinear(c,p)=1$ (Proposition~\ref{pr:mol.special}), so $\partial \maximinOptimalLinear(c,p)/\partial c = 0$.
  If $c > p/(1-p)$, then $\maximinOptimalLinear(c,p)\in(0,1)$ and, as an interior maximizer, satisfies the first-order optimality condition
  \[
    \frac{\partial \underline{\Gamma}(c,p,s)}{\partial s}\Bigg|_{s=\maximinOptimalLinear(c,p)}
    = 0.
  \]

  Next, we show that $f(c,p,s) > 0$ for all $c>0$, $p\in (0,1)$, and $s\in [0,1]$.
  For $s<1$, we have
  \begin{align*}
    f(c,p,s)
    &= \frac{p}{2(1+pc)}-\frac{1}{2c}\sum_{i=0}^\infty p(1-p)^i \frac{cs(1-s)^i}{1+cs(1-s)^i}\\
    &\gevar{(a)} \frac{p}{2(1+pc)}-\frac{\hat{\mu}}{2c(1+\hat{\mu})}
    > 0,
  \end{align*}
  where (a) follows from Jensen's inequality applied to the concave function $x/(1+x)$ on $x\ge 0$, and where
  \begin{align*}
    \hat{\mu}
    &\eqdef psc \sum_{i=0}^\infty [(1-p)(1-s)]^i
    = \frac{c}{1/p+1/s-1}
    < pc.
  \end{align*}
  For $s=1$, we also have
  \[
    f(c,p,1)
    = \frac{p}{2(1+pc)}-\frac{p}{2(1+c)}
    > 0.
  \]
  Therefore, $\overline{G}(c,p,s)$ is strictly increasing on $c>0$, whether $s$ is fixed or $s=\maximinOptimalLinear(c,p)$.

  Furthermore, for fixed $s\in (0,1)$,
  \begin{align*}
    \lim_{c\to +\infty} \overline{G}(c,p,s)
    &= \lim_{c\to +\infty} \left(\frac{1}{2}\log(1+pc) - \sum_{i=0}^\infty p(1-p)^i \frac{1}{2}\log(1+cs(1-s)^i)\right)\\
    &= \lim_{c\to +\infty} \frac{1}{2} \sum_{i=0}^\infty p(1-p)^i \log\frac{1+pc}{1+cs(1-s)^i}\\
    &= \frac{1}{2} \sum_{i=0}^\infty p(1-p)^i \log\frac{p}{s(1-s)^i}\\
    &= \frac{1}{2}\left(\log\frac{p}{s} - \frac{1-p}{p}\log(1-s)\right)
    = G_p(s).
  \end{align*}
  This equation also holds for $s=0$ and $s=1$, with $\lim_{c\to +\infty} \overline{G}(c,p,s) = G_p(s) = +\infty$.
\end{proofof}

\begin{proofof}{Theorem~\ref{th:ag_minimax}}
  From Proposition~\ref{pr:ag_increasing_in_c}, we have
  \[
    \min_{s\in [0,1]} \sup_{c>0} \overline{G}(c,p,s)
    = \min_{s\in [0,1]} \lim_{c\to +\infty} \overline{G}(c,p,s)
    = \min_{s\in [0,1]} G_p(s)
  \]
  and
  \[
    \sup_{c>0} \min_{s\in [0,1]} \overline{G}(c,p,s)
    = \lim_{c\to +\infty} \overline{G}(c,p,\maximinOptimalLinear(c,p)).
  \]
  It is easy to verify that $G_p(s)$ is strictly convex in $s$ and that its unique minimizer is $s=p$.
  Thus, $\caagOptimalLinear(p)=p$ and
  \[
    \min_{s\in [0,1]} \sup_{c>0} \overline{G}(c,p,s)
    = G_p(p)
    \ge \sup_{c>0} \min_{s\in [0,1]} \overline{G}(c,p,s).
  \]
  Moreover, since the worst-case throughput of the maximin optimal linear policy $\maximinOptimalLinear(c,p)$ is bounded above by the worst-case throughput of the maximin optimal policy $\maximinOptimalPolicy[p]$, it follows that
  \[
    \sup_{c>0} \min_{s\in [0,1]} \overline{G}(c,p,s)
    \ge \sup_{c>0} \overline{G}(c,p,\maximinOptimalPolicy[p])
    = G_p(p) \quad \text{(\cite[Thm.~3.29]{yang2025power})},
  \]
  where we write $\overline{G}(c,p,\maximinOptimalPolicy[p])$ for the nominal additive gap of $\maximinOptimalPolicy[p]$ (by abuse of notation).

  Therefore,
  \[
    \min_{s\in [0,1]} \sup_{c>0} \overline{G}(c,p,s)
    = \sup_{c>0} \min_{s\in [0,1]} \overline{G}(c,p,s)
    = \lim_{c\to +\infty} \overline{G}(c,p,\maximinOptimalLinear(c,p)) = G_p(p).
  \]
  Hence, $\lim_{c\to +\infty} \maximinOptimalLinear(c,p) = p = \caagOptimalLinear(p)$; otherwise, $\liminf_{c\to +\infty} \overline{G}(c,p,\maximinOptimalLinear(c,p)) < G_p(p)$, a contradiction.
\end{proofof}

\begin{proofof}{Theorem~\ref{th:mf_minimax}}
  First note that if Conjecture~\ref{cj:throughput_quasi_concave_in_s} holds, then $\overline{F}(c,p,s)$ is strictly quasiconcave in $s$ for $p\in (0,1)$ and $c>0$.
  If Conjecture~\ref{cj:mf_quasi_convex_in_c} also holds, then by~\cite[Thm.~3]{simons2009minimax},
  \[
    \max_{s\in [0,1]} \inf_{c>0} \underline{F}(c,p,s)
    = \inf_{c>0} \max_{s\in [0,1]} \underline{F}(c,p,s).
  \]

  Next, by Theorem~\ref{th:ag_minimax},
  \begin{align*}
    \lim_{c\to +\infty} \max_{s\in [0,1]} \underline{F}(c,p,s)
    &= \lim_{c\to +\infty} \left( 1-\frac{\overline{G}(c,p,\maximinOptimalLinear(c,p))}{r(pc)} \right)
    = 1 - \frac{G_p(p)}{\lim_{c\to +\infty} r(pc)}
    = 1.
  \end{align*}
  On the other hand, for $c=p/(1-p)$, it follows from Proposition~\ref{pr:mol.special} that
  \[
    \max_{s\in [0,1]} \underline{F}(c,p,s)
    = \frac{\underline{\Gamma}(c,s,1)}{r(pc)}
    = \frac{pr(c)}{r(pc)}
    < 1.
  \]
  This implies that $c=+\infty$ cannot be a ``minimizer'' of $\max_{s\in [0,1]} \underline{F}(c,p,s)$ in $c$, so a finite minimizer $c_{\times}(p)$ exists.
  Consequently, $(c_{\times}(p),\camfOptimalLinear(p))$ is a saddle point and
  \[
    \underline{F}(c_{\times}(p),p,\camfOptimalLinear(p))
    = \max_{s\in [0,1]} \underline{F}(c_{\times}(p),p,s)
    = \underline{F}(c_{\times}(p),p,\maximinOptimalLinear(c_{\times}(p),p)).
  \]
  By the strict quasiconcavity of $\underline{F}(c,p,s)$ in $s$, we have $\camfOptimalLinear(p) = \maximinOptimalLinear(c_{\times}(p),p)$.
  If there were another saddle point $(\tilde{c},\tilde{s})$ (necessarily with $\tilde{s}>0$), then both $(\tilde{c},\camfOptimalLinear(p))$ and $(c_{\times}(p),\tilde{s})$ would also be saddle points.
  The strict quasiconcavity in $s$ together with the strict quasiconvexity in $c$ forces $\tilde c=c_{\times}(p)$ and $\tilde s=\camfOptimalLinear(p)$, establishing uniqueness of the saddle point and of the minimizer \(c_{\times}(p)\).
\end{proofof}

\begin{proofof}{Lemma~\ref{le:cs_asymptotic}}
  On the one hand,
  \begin{align*}
    \underline{\Gamma}(b/p, p, ap)
    &= \sum_{i=0}^\infty p(1-p)^{i} r(ab(1-ap)^{i})
    \levar{(a)} \sum_{i=0}^\infty p\en^{-pi} r(ab\en^{-api})\\
    &\levar{(b)} \sum_{i=0}^{\infty} \left( \int_{pi}^{p(i+1)} \en^{-x} r(ab\en^{-ax}) \diff x + \frac{p^2r(ab)}{2} \en^{-pi} + \frac{ap^2}{4}\min\{ab\en^{-(a+1)pi}, \en^{-pi}\}\right)\\
    &= \int_0^{+\infty} \en^{-x} r(ab\en^{-ax}) \diff x + \frac{p^2r(ab)}{2(1-\en^{-p})} + \frac{ap^2}{4}\min\left\{\frac{ab}{1-\en^{-(a+1)p}}, \frac{1}{1-\en^{-p}}\right\}\\
    &\levar{(c)} \int_0^{+\infty} \en^{-x} r(ab\en^{-ax}) \diff x + pr(ab) + \frac{ap}{2}\min\{ab, b, 1\},
  \end{align*}
  where (a) follows from $1+x\le \en^x$ for $x\in\real$, (b) from Lemma~\ref{integration_bound} with
  \begin{align*}
    \sup_{x\in [pi,p(i+1)]} |(\en^{-x} r(ab\en^{-ax}))'|
    &= \sup_{x\in [pi,p(i+1)]} (\en^{-x} r(ab\en^{-ax})+a^2b\en^{-(1+a)x}r'(ab\en^{-ax}))\\
    &\le \en^{-pi}r(ab) + \sup_{x\in [pi,p(i+1)]} \frac{a^2b\en^{-x}}{2(\en^{ax}+ab)}\\
    &\le \en^{-pi}r(ab) + \frac{a}{2}\min\{ab\en^{-(a+1)pi}, \en^{-pi}\},
  \end{align*}
  and (c) from $1-\en^{-x}\ge x/2$ for $x\in (0,1)$.

  On the other hand,
  \begin{align*}
    \underline{\Gamma}(b/p, p, ap)
    &\gevar{(a)} \sum_{i=0}^\infty p\en^{-pi/(1-p)} r(ab\en^{-api/(1-ap)})\\
    &\gevar{(b)} (1-a'p) \sum_{i=0}^\infty \int_{pi/(1-a'p)}^{p(i+1)/(1-a'p)}\en^{-pi/(1-a'p)} r(ab\en^{-api/(1-a'p)}) \diff x\\
    &= (1-a'p) \int_0^{+\infty} \en^{-x} r(ab\en^{-ax}) \diff x
    \ge \int_0^{+\infty} \en^{-x} r(ab\en^{-ax}) \diff x - a'pr(ab),
  \end{align*}
  where $a'=\max\{a,1\}$, (a) follows from $1+x\ge \en^{x/(1+x)}$ for $x>-1$, and (b) from Lemma~\ref{integration_bound}.
\end{proofof}

\begin{proofof}{Theorem~\ref{th:worst_mf}}
  If Conjecture~\ref{cj:worst_mf_at_zero} holds, then
  \[
    \inf_{p\in (0,1)} F_p
    = \lim_{p\to 0} \underline{F}(c_{\times}(p),p,\camfOptimalLinear(p)).
  \]
  Define $a_{\times}(p)\eqdef \camfOptimalLinear(p) / p$ and $b_{\times}(p)\eqdef pc_{\times}(p)$.
  By Lemma~\ref{le:cs_asymptotic}, as $p\to 0$,
  \begin{equation}
    - \mathrm{O}(p\max\{a,1\}r(ab))
    \le \underline{\Gamma}(b/p,p,ap) - \Gamma_0(a,b)
    \le \mathrm{O}(p(a+r(ab))).\label{eq:worst_mf_proof.asymptotic}
  \end{equation}
  Thus,
  \begin{equation}
    \lim_{p\to 0} \underline{F}(c_{\times}(p),p,\camfOptimalLinear(p))
    = \lim_{p\to 0} \frac{\Gamma_0(a_{\times}(p),b_{\times}(p))}{r(b_{\times}(p))}
    = \frac{\Gamma_0(\tilde{a},\tilde{b})}{r(\tilde{b})},\label{eq:worst_mf_proof.limit}
  \end{equation}
  where $\tilde{a}\eqdef \lim_{p\to 0} a_{\times}(p)$ and $\tilde{b}\eqdef \lim_{p\to 0} b_{\times}(p)$.

  Next, we show that
  \[
    \tilde{b}
    = \arg \inf_{b>0} \frac{\Gamma_0(\tilde{a},b)}{r(b)}.
  \]
  If not, then there exists $b' \neq \tilde{b}$ such that $\rho' \eqdef \Gamma_0(\tilde{a},b')/r(b') < \tilde{\rho} \eqdef \Gamma_0(\tilde{a},\tilde{b})/r(\tilde{b})$.
  Let $\epsilon \eqdef (\tilde{\rho} - \rho') / 3$.
  By~\eqref{eq:worst_mf_proof.asymptotic}, there exists $\delta_1 > 0$ such that for all $p\in (0,\delta_1)$,
  \[
    \underline{F}(b'/p, p, \camfOptimalLinear(p))
    \le \rho' + \epsilon.
  \]
  By~\eqref{eq:worst_mf_proof.limit}, there exists $\delta_2 > 0$ such that for all $p\in (0,\delta_2)$,
  \[
    \underline{F}(c_{\times}(p),p,\camfOptimalLinear(p))
    \ge \tilde{\rho} - \epsilon.
  \]
  Then for all $p\in (0,\min\{\delta_1,\delta_2\})$, we have
  \[
    \underline{F}(b'/p, p, \camfOptimalLinear(p))
    \le \rho' + \epsilon
    = \tilde{\rho} - 2\epsilon
    \le \underline{F}(c_{\times}(p),p,\camfOptimalLinear(p)) - \epsilon,
  \]
  which contradicts the definition of $c_{\times}(p)$.

  Finally, we show that
  \[
    \tilde{a}
    = \arg \sup_{a>0} \inf_{b>0} \frac{\Gamma_0(a,b)}{r(b)}.
  \]
  If not, then there exists $a' \neq \tilde{a}$ such that $\rho' \eqdef \inf_{b>0} \Gamma_0(a',b)/r(b) > \hat{\rho} \eqdef \inf_{b>0} \Gamma_0(\tilde{a},b)/r(b)$.
  Let $\epsilon \eqdef (\rho' - \hat{\rho}) / 3$.
  By~\eqref{eq:worst_mf_proof.asymptotic},
  \begin{align*}
    \inf_{c>0} \underline{F}(c,p,a'p)
    &= \inf_{b>0} \underline{F}(b/p,p,a'p)
    \ge \inf_{b>0} \left(\frac{\Gamma_0(a',b)}{r(b)} - \frac{\mathrm{O}(p\max\{a',1\}r(a'b))}{r(b)}\right) \\
    &= \inf_{b>0} \frac{\Gamma_0(a',b)}{r(b)} - \mathrm{O}(p(\max\{a',1\})^2),
  \end{align*}
  so there exists $\delta_1 > 0$ such that for all $p\in (0,\delta_1)$,
  \[
    \inf_{c>0} \underline{F}(c,p,a'p)
    \ge \rho' - \epsilon.
  \]
  By~\eqref{eq:worst_mf_proof.limit}, there exists $\delta_2 > 0$ such that for all $p\in (0,\delta_2)$,
  \begin{align*}
    \inf_{c>0} \underline{F}(c,p,\camfOptimalLinear(p))
    &= \underline{F}(c_{\times}(p),p,\camfOptimalLinear(p))
    \le \frac{\Gamma_0(\tilde{a},\tilde{b})}{r(\tilde{b})} + \epsilon \\
    &= \inf_{b>0} \frac{\Gamma_0(\tilde{a},b)}{r(b)} + \epsilon
    = \hat{\rho} + \epsilon.
  \end{align*}
  Then for all $p\in (0, \min\{\delta_1,\delta_2\})$, we have
  \[
    \inf_{c>0} \underline{F}(c,p,a'p)
    \ge \rho' - \epsilon
    = \hat{\rho} + 2\epsilon
    \ge \inf_{c>0} \underline{F}(c,p,\camfOptimalLinear(p)) + \epsilon,
  \]
  which contradicts the definition of $\camfOptimalLinear(p)$.
\end{proofof}

\section{Proofs of Results in Section~\ref{subsec:greedy_policy_optimality}}\label{sec:proofs_of_greedy_policy_optimality}

\begin{proofof}{Lemma~\ref{f_bound_to_c_bound}}
  By definition, for any $c < \underline{c}'(\underline{x},\overline{x},\mu)$,
  \[
    r'(c)
    \ge \overline{f}(c,\underline{x},\overline{x},\mu)
    \ge \int_{[0,c)} r'(x) \diff Q
  \]
  for all $Q\in Q'_{\underline{x},\overline{x},\mu}$, which implies $c \le c^*(Q)$ for all $Q\in Q'_{\underline{x},\overline{x},\mu}$, hence $c \le \underline{c}(\underline{x},\overline{x},\mu)$, and therefore $\underline{c}'(\underline{x},\overline{x},\mu) \le \underline{c}(\underline{x},\overline{x},\mu)$.
  On the other hand, for any $c < \underline{c}(\underline{x},\overline{x},\mu)$,
  \[
    r'(c)
    \ge \int_{[0,c)} r'(x) \diff Q
  \]
  for all $Q\in Q'_{\underline{x},\overline{x},\mu}$, that is, $r'(c) \ge \overline{f}(c,\underline{x},\overline{x},\mu)$, which implies $c\le \underline{c}'(\underline{x},\overline{x},\mu)$, and hence $\underline{c}(\underline{x},\overline{x},\mu)\le \underline{c}'(\underline{x},\overline{x},\mu)$.
  Therefore, $\underline{c}(\underline{x},\overline{x},\mu) = \underline{c}(\underline{x},\overline{x},\mu)$.

  Similarly, for any $c < \overline{c}'(\underline{x},\overline{x},\mu)$,
  \[
    r'(c)
    > r'\left(\frac{c+\overline{c}'(\underline{x},\overline{x},\mu)}{2}\right)\\
    \ge \underline{f}\left(\frac{c+\overline{c}'(\underline{x},\overline{x},\mu)}{2},\underline{x},\overline{x},\mu\right)
    \ge \underline{f}(c,\underline{x},\overline{x},\mu),
  \]
  and hence
  \[
    r'(c)\ge \int_{[0,c)} r'(x) \diff Q
  \]
  for some $Q\in Q'_{\underline{x},\overline{x},\mu}$.
  This implies $c \le c^*(Q) \le \overline{c}(\underline{x},\overline{x},\mu)$, and hence $\overline{c}'(\underline{x},\overline{x},\mu) \le \overline{c}(\underline{x},\overline{x},\mu)$.
  Moreover, for any $c < \overline{c}(\underline{x},\overline{x},\mu)$, there exists a $Q\in Q'_{\underline{x},\overline{x},\mu}$ such that
  \[
    r'(c)
    \ge \int_{[0,c)} r'(x) \diff Q
    \ge \underline{f}(c,\underline{x},\overline{x},\mu).
  \]
  This implies $c\le \overline{c}'(\underline{x},\overline{x},\mu)$, and hence $\overline{c}(\underline{x},\overline{x},\mu)\le \overline{c}'(\underline{x},\overline{x},\mu)$.
\end{proofof}

\begin{proofof}{Lemma~\ref{f_bounds}}
  The problem to be solved is a linear program in a measure space.
  By \cite[Thm.~3.1]{lai_linear_1994}, the optimal value, minimum or maximum, must occur at an extreme point of the set of feasible probability measures, all probability measures $Q$ satisfying the constraints
  \[
    \int_{\nnreal} \diff Q = 1
    \
    \text{and}
    \
    \int_{\nnreal} x\diff Q = \mu.
  \]
  It follows from \cite[Thm.~3.2]{lai_linear_1994} that such an extreme point $Q$ must be a discrete probability measure concentrated at one or two points.
  Therefore, it suffices to consider $Q$ of the form
  \(
  Q(\mu)=1
  \)
  or
  \[
    Q(a) = \frac{b - \mu}{b-a}\ \text{and}\ Q(b) = \frac{\mu - a}{b - a}
  \]
  with $\underline{x}\le a<\mu<b\le \overline{x}$.
  The optimization problem then reduces to the following simplified forms:
  \[
    \underline{g}(c,\underline{x},\overline{x},\mu)
    \eqdef \begin{cases}
             0 &$c \le \mu$,\\
             \displaystyle\min\left\{\underline{g}_{a,b}(c), \frac{1}{1+\mu}\right\} &$c > \mu$,
    \end{cases}
  \]
  and
  \[
    \overline{g}(c,\underline{x},\overline{x},\mu)
    \eqdef \begin{cases}
             \overline{g}_{a,b}(c) &$c \le \mu$,\\[1ex]
             \displaystyle\max\left\{\overline{g}_{a,b}(c), \frac{1}{1+\mu}\right\} &$c > \mu$,
    \end{cases}
  \]
  where
  \begin{gather*}
    \underline{g}_{a,b}(c)
    \eqdef \inf_{\underline{x}\le a<\mu<b\le \overline{x}} g_{a,b}(c),\\
    \overline{g}_{a,b}(c)
    \eqdef \sup_{\underline{x}\le a<\mu<b\le \overline{x}} g_{a,b}(c),\\
    g_{a,b}(c)
    \eqdef \frac{b-\mu}{(b-a)(1+a)}1\{c>a\} + \frac{\mu -a}{(b-a)(1+b)}1\{c> b\}.
  \end{gather*}

  1) If $c>b$, then
  \[
    g_{a,b}(c)
    = \frac{b-\mu}{(b-a)(1+a)} + \frac{\mu -a}{(b-a)(1+b)}.
  \]
  By the convexity of $1/(1+x)$ (for $x\ge 0$), the infimum and the supremum of $g_{a,b}(c)$ are attained as $(a,b)\to (\mu,\mu)$ (Jensen's inequality) and $(a,b)\to (\underline{x},c')$ (\cite[Lemma~2]{shaviv2016universally}), respectively, where $c'\eqdef \min\{c,\overline{x}\}$, so
  \[
    \underline{g}_{a,b}(c) = \frac{1}{1+\mu}
    \
    \text{and}
    \
    \overline{g}_{a,b}(c) = \frac{1+\underline{x}+c'-\mu}{(1+\underline{x})(1+c')}.
  \]

  2) If $\mu<c\le b$, then
  \begin{equation}
    g_{a,b}(c)
    = \frac{b-\mu}{(b-a)(1+a)}
    = \frac{1}{1+a} - \frac{\mu-a}{(b-a)(1+a)}, \label{a_c_b}
  \end{equation}
  which is strictly increasing in $b$ for any fixed $a$ and $c$.
  On the other hand,
  \[
    g_{a,b}(c)
    = \frac{b-\mu}{-\left(a-\frac{b-1}{2}\right)^2+\frac{(b+1)^2}{4}},
  \]
  which is strictly decreasing and strictly increasing in $a$ for $a<(b-1)/2$ and $a>(b-1)/2$, respectively.
  Thus, taking $b=c$ and according to the position of
  \[
    a_0
    \eqdef \frac{b-1}{2}
    = \frac{c-1}{2}
  \]
  (compared to $\underline{x}$ and $\mu$), we have
  \begin{align*}
    \underline{g}_{a,b}(c)
    &= \begin{cases}
         g_{\underline{x},c}(c) &$a_0<\underline{x}$,\\
         g_{a_0,c}(c) &$\underline{x}\le a_0<\mu$,\\
         g_{\mu,c}(c) &$a_0\ge \mu$,
    \end{cases}
    = \begin{cases}
        \frac{c-\mu}{(c-\underline{x})(1+\underline{x})} &$c\in [0,\iota(\underline{x}))\cap (\mu, \overline{x}]$,\\
        \frac{4(c-\mu)}{(c+1)^2} &$c\in [\iota(\underline{x}),\iota(\mu))\cap (\mu, \overline{x}]$,\\
        \frac{1}{1+\mu} &$c\in [\iota(\mu),+\infty)\cap (\mu, \overline{x}]$.
    \end{cases}
  \end{align*}
  Similarly, taking $b=\overline{x}$ and comparing
  \begin{equation}
    a_1
    \eqdef \frac{b-1}{2}
    = \frac{\overline{x}-1}{2} \label{a_1}
  \end{equation}
  with $a_2\eqdef (\underline{x}+\mu)/2$, we have
  \begin{align*}
    \overline{g}_{a,b}(c)
    &= \begin{cases}
         g_{\mu,\overline{x}}(c) &$a_1<a_2$,\\
         g_{\underline{x},\overline{x}}(c) &$a_1\ge a_2$,
    \end{cases}
    = \begin{cases}
        \frac{1}{1+\mu} &$\mu>\tau$,\\
        \frac{\overline{x}-\mu}{(\overline{x}-\underline{x})(1+\underline{x})} &$\mu\le\tau$.
    \end{cases}
  \end{align*}

  3) If $a<c\le\mu$, we also have \eqref{a_c_b}.
  Thus, taking $b=\overline{x}$ and comparing $a_1$ (defined by \eqref{a_1}) with $a_3\eqdef (\underline{x}+c)/2$, we have
  \begin{align*}
    \overline{g}_{a,b}(c)
    &= \begin{cases}
         g_{c,\overline{x}}(c) &$a_1<a_3$,\\
         g_{\underline{x},\overline{x}}(c) &$a_1\ge a_3$,
    \end{cases}
    = \begin{cases}
        \frac{\overline{x}-\mu}{(\overline{x}-c)(1+c)} &$c>\tau$,\\
        \frac{\overline{x}-\mu}{(\overline{x}-\underline{x})(1+\underline{x})} &$c\le\tau$.
    \end{cases}
  \end{align*}

  4) If $c\le a$, then $g_{a,b}(c)=0$.

  Combining Parts (1)--(4) gives \eqref{f_bounds.lower} and \eqref{f_bounds.upper}.
  Some of the cases in \eqref{f_bounds.upper} are slightly complicated, because the comparison of several candidates of the maximum are need as follows:
  \[
    \frac{1}{1+\mu}
    < \frac{1+\underline{x}+c-\mu}{(1+\underline{x})(1+c)}
    \quad \text{for $c\in (\mu,\overline{x}]$}
  \]
  and
  \[
    \frac{\overline{x}-\mu}{(\overline{x}-\underline{x})(1+\underline{x})}
    \lesseqqgtr \frac{1+\underline{x}+c-\mu}{(1+\underline{x})(1+c)}
    \quad \text{for $c\gtreqqless \tau$ and $c\in (\mu,\overline{x}]$.}
  \]
\end{proofof}

\begin{proofof}{Theorem~\ref{thm:grdy:lower_c}}
  It is clear that
  \[
    \underline{c}_1
    \lesseqqgtr \tau \quad \text{for $\mu\lesseqqgtr\tau$}.
  \]
  By Lemmas~\ref{f_bound_to_c_bound} and \ref{f_bounds},
  \begin{align*}
    \underline{c}(\underline{x},\overline{x},\mu)
    &= \sup([0,\underline{x}] \cup A \cup B)
    = \begin{cases}
        \underline{c}_1 &$\mu<\tau$,\\
        \mu &$\mu\ge\tau$,
    \end{cases}
  \end{align*}
  where
  \begin{align*}
    A
    &= [0, \underline{c}_1] \cap [0,\tau] \cap (\underline{x},\overline{x}]
    = \begin{cases}
        (\underline{x},\underline{c}_1] &$\mu<\tau$,\\{}
        (\underline{x},\tau] &$\mu\ge\tau$,
    \end{cases}\\
    B
    &= (\tau,+\infty)\cap (\underline{x},\mu].
  \end{align*}
\end{proofof}

\begin{proofof}{Theorem~\ref{thm:grdy:upper_c}}
  Note that
  \[
    \overline{c}_1
    \lesseqqgtr \iota(\underline{x})
    \quad \text{for $\mu\lesseqqgtr\frac{3}{2}\underline{x}+\frac{1}{2}$}
  \]
  and
  \[
    \overline{c}_2
    \lesseqqgtr \iota(\underline{x})
    \quad \text{for $\mu\lesseqqgtr\frac{3}{2}\underline{x}+\frac{1}{2}$}.
  \]
  By Lemmas~\ref{f_bound_to_c_bound} and \ref{f_bounds},
  \begin{align*}
    \overline{c}(\underline{x},\overline{x},\mu)
    &= \sup([0,\mu] \cup A \cup B)
    = \begin{cases}
        \min\{\overline{c}_1,\overline{x}\} &$\mu<\frac{3}{2}\underline{x}+\frac{1}{2}$,\\
        \min\{\overline{c}_2,\overline{x}\} &$\mu\ge\frac{3}{2}\underline{x}+\frac{1}{2}$,
    \end{cases}
  \end{align*}
  where
  \begin{gather*}
    A
    = [0, \overline{c}_1] \cap [0, \iota(\underline{x})] \cap (\mu, \overline{x}],\\
    B
    = [\iota(\underline{x}),\overline{c}_2] \cap (\mu,\overline{x}].
  \end{gather*}
\end{proofof}

\begin{proofof}{Theorem~\ref{upper_c_for_mcr}}
  Let $f(x)\eqdef \overline{c}'(0,x)$.
  By Theorem~\ref{lower_upper_c_for_clipped},
  \[
    f(x)
    = \begin{cases}
        \frac{x+\sqrt{x^2+4x}}{2} &$x<\frac{1}{2}$,\\
        \frac{4}{3}x+\frac{1}{3} &$x\ge\frac{1}{2}$.
    \end{cases}
  \]
  Then for $x<\frac{1}{2}$,
  \[
    f'(x)
    = \frac{1}{2}+\frac{x+2}{2(x^2+4x)^{1/2}}
    > 0
  \]
  and
  \begin{align*}
    f''(x)
    &= \frac{2(x^2+4x)^{1/2}-2(x+2)^2(x^2+4x)^{-1/2}}{4(x^2+4x)}\\
    &= \frac{2(x^2+4x)-2(x+2)^2}{4(x^2+4x)^{3/2}}
    = -\frac{2}{(x^2+4x)^{3/2}}
    < 0.
  \end{align*}
  It is clear that $f'(0)=+\infty$ and $f$ is differentiable at $x=1/2$ with $f'(\frac{1}{2})=\frac{4}{3}$.
  Hence $f$ is strictly increasing and concave on $[0,+\infty)$, and therefore, for every $p\in (0,1)$, $\overline{c}''(p)$ is the unique positive solution of $c=f(pc)$ (if exists) or $+\infty$.
  Solving the equation then gives
  \[
    \overline{c}''(p)
    = \begin{cases}
        \frac{p}{1-p} &$p\in (0,\frac{1}{2})$,\\
        \frac{1}{3-4p} &$p\in [\frac{1}{2},\frac{3}{4})$,\\
        +\infty &$p\in [\frac{3}{4},1)$.
    \end{cases}
  \]
  The verification of the remaining part of the theorem is straightforward.
\end{proofof}

\section{Numerical Verification of Conjectures in Sec.~\ref{subsec:saddle_point_structures}}
\label{sec:numerical_verification}

To provide robust numerical evidence for the claims, we employ the following methods in our numerical verification.

\begin{enumerate}
  \item The verification typically involves evaluating a one-variable function $f_{\xi_1,\ldots,\xi_k}(x)$ with parameters $(\xi_1, \ldots, \xi_k)$.
  When $c$, $p$, or $s$ is the varying parameter, we numerate them on the following default grids:
  \begin{gather*}
    A
    \eqdef \{j\times 10^i: -3\le i\le 2, 1\le j\le 9\}\cup \{10^3\},\\
    B
    \eqdef \{10^{-3}\} \cup \{0.01 i: 1\le i\le 99\},\\
    C
    \eqdef B \cup \{1\},
  \end{gather*}
  used respectively for $c$, $p$, and $s$.

  \item To verify qualitative properties of a function $f(x)$, we use adaptive sampling so that each pair of adjacent samples $(x,f(x))$ and $(x',f(x'))$ satisfies
  \(
  |x-x'|\le d_1
  \)
  or
  \[
    \sqrt{(x-x')^2+(f(x)-f(x'))^2}\le d_2;
  \]
  we set \(d_1=10^{-4}\) and \(d_2=10^{-3}\).

  \item In some cases, the domain of a function is not bounded.
  For example, the range of $c$ is $[0,+\infty)$.
  In this case, we consider a monotone transform, e.g., $g(c')\eqdef c'/(1-c')$.
  Then, the domain of the new function $\hat{f}(c')\eqdef f(g(c'))$ is $[0,1)$.
\end{enumerate}

\begin{verificationOf}{Conjecture~\ref{cj:throughput_quasi_concave_in_s}}
  %eh4-4
  For fixed $(c,p)\in A\times B$, the worst-case throughput $\underline{\Gamma}(c,p,s)$ is strictly quasiconcave in $s$.
\end{verificationOf}

\begin{verificationOf}{Conjecture~\ref{cj:mf_quasi_convex_in_c}}
  % eh4-14
  For fixed $(p,s)\in B\times C$, the nominal multiplicative factor $\underline{F}(c,p,s)$ is strictly quasiconvex in $c$.
\end{verificationOf}

\begin{verificationOf}{Conjecture~\ref{cj:worst_mf_at_zero}}
  % eh4-19
  Numerically, $\inf_{p\in B} F_p = F_{0.001} \approx 0.6532$.
  Table~\ref{tab:mf_infimum_verification} reports the convergence of $F_p$, $pc_{\times}(p)$, and $\camfOptimalLinear(p)/p$ as $p\to 0$; the observed limits agree with Theorem~\ref{th:worst_mf} under Conjecture~\ref{cj:worst_mf_at_zero}.
  \begin{table}
    \caption{The Convergence of $F_p$, $pc_{\times}(p)$, and $\camfOptimalLinear(p)/p$ as $p\to 0$}
    \label{tab:mf_infimum_verification}
    \centering
    \begin{tabular}{cccccc}
      \toprule
      $p$     & $c_\times(p)$ & $\camfOptimalLinear(p)$ & $F_p$    & $pc_{\times}(p)$ & $\camfOptimalLinear(p)/p$ \\
      \midrule
      0.10000 & 19.712069     & 0.205705                & 0.674155 & 1.971207         & 2.057054                  \\
      0.01000 & 181.016019    & 0.022600                & 0.655090 & 1.810160         & 2.260028                  \\
      0.00100 & 1795.415833   & 0.002282                & 0.653247 & 1.795416         & 2.282255                  \\
      0.00010 & 17939.541832  & 0.000228                & 0.653063 & 1.793954         & 2.284499                  \\
      0.00001 & 179380.373113 & 0.000023                & 0.653045 & 1.793804         & 2.284723                  \\
      \bottomrule
    \end{tabular}
  \end{table}
\end{verificationOf}

\section{Auxiliary Results}

\begin{lemma}\label{integration_bound}
  If $f: [a,b]\to\real$ is continuous on $[a,b]$, differentiable on $(a,b)$, and satisfies $f(x)\le f(a)$ for all $x\in [a,b]$, then
  \[
    f(a)(b-a)-\frac{1}{2} A (b-a)^2
    \le \int_a^b f(x) \diff x
    \le f(a)(b-a),
  \]
  where $A=\sup_{x\in (a,b)} |f'(x)|$.
\end{lemma}

\begin{IEEEproof}
  Observe that
  \(
  0
  \le f(a)-f(x)
  \le A(x-a)
  \)
  by the mean value theorem,
  Taking integration over $[a,b]$, we obtain
  \[
    0
    \le \int_a^b (f(a)-f(x)) \diff x
    \le \frac{1}{2} A(b-a)^2,
  \]
  which concludes the lemma.
\end{IEEEproof}

\bibliographystyle{IEEEtran}
\bibliography{IEEEabrv,eh4}

\end{document}